\documentclass[
 aip,
 amsmath,amssymb,
 preprint
]{revtex4-2}

\usepackage{physics}
\usepackage{graphicx}
\usepackage{verbatim}
\usepackage{hyperref}
\usepackage{natbib}
\usepackage{placeins}
\usepackage{adjustbox}
\usepackage{soul}
\usepackage{todonotes}
\usepackage{tikz}
\usepackage{subcaption}
\usepackage{cleveref}
\usepackage{float}
\usepackage{subfloat}
\usepackage{color, colortbl}
\definecolor{LightCyan}{rgb}{0.88,1,1}

\Crefname{equation}{Eq.}{Eqs.}
\Crefname{figure}{Fig.}{Figs.}
\DeclareMathAlphabet{\mathpzc}{OT1}{pzc}{m}{it}
\DeclareMathAlphabet\mathbfcal{OMS}{cmsy}{b}{n}

\begin{document}
	
	\title{The role of  noise in PIC and Vlasov simulations of the Buneman instability}
	
	\author{Arash Tavassoli}
	\email{art562@usask.ca}
    \author{Oleksandr Chapurin}
 	\author{Marilyn Jimenez}
    \author{Mina Papahn Zadeh}
    \author{Trevor Zintel}
    \author{Meghraj Sengupta}
    \author{L\'ena\"ic Cou\"edel}
	\affiliation{Department of Physics and Engineering Physics, University of Saskatchewan, 116 Science Place, Saskatoon SK, S7N 5E2 CANADA}
    \author{Raymond J. Spiteri}
	\affiliation{Department of Computer Science, University of Saskatchewan,  Saskatoon, Saskatchewan, Canada S7N 5C9}
	 \author{Magdi Shoucri}
	 \author{Andrei Smolyakov}
	\affiliation{Department of Physics and Engineering Physics, University of Saskatchewan, 116 Science Place, Saskatoon SK, S7N 5E2 CANADA}

	\date{\today}

\begin{abstract}
The effects of noise in particle-in-cell (PIC) and Vlasov simulations of the Buneman instability in unmagnetized plasmas are studied. It is found that, in the regime of low drift velocity, the linear stage of the instability in PIC simulations differs significantly from the theoretical predictions, whereas in the Vlasov simulations it does not.  A series of highly resolved PIC simulations with increasingly large numbers of macroparticles per cell is performed using a number of different PIC codes. 
 All the simulations predict highly similar growth rates that are several times larger than those calculated from the linear theory. As a result, we find that the true convergence of the PIC simulations in the linear regime is elusive to achieve in practice and can easily be misidentified.
The discrepancy between the theoretical and observed growth rates is attributed to the initial noise inherently present in PIC simulations, but not in Vlasov simulations, that causes particle trapping even though the fraction of trapped particles is low.  We show analytically that even weak distortions of the electron velocity distribution function (such as flattening due to particle trapping) result in significant modifications of the growth rates. It is also found that the common quiet-start method for PIC simulations leads to more accurate growth rates but only if the maximum growth rate mode is perturbed initially. We demonstrate that the quiet-start method does not completely remedy the noise problem because the simulations generally exhibit inconsistencies with the linear theory. 
\end{abstract}

\maketitle
\section{Introduction}

Kinetic simulations are a powerful tool for studies of the linear and nonlinear behavior of plasmas.  The particle-in-cell (PIC) method and the continuum Vlasov method are two widely used simulation approaches. The PIC method, which has been available for several decades, has successfully captured many physical phenomena in various scenarios.  The PIC method, however, is also known for relatively large levels of numerical noise introduced  by the discretization and the limited number of macroparticles used to resolve the phase space\cite{langdon1979kinetic}. The noise in PIC simulations exists during the first time step (initial noise), but it also evolves during simulations, thus affecting the results.
An alternative to the PIC method, the continuum Vlasov method,  is well known as a method that is free of statistical noise.  The availability of high-performance computational resources has led to a steady increase of interest and applications of continuum simulations for many physical phenomena and situations that are poorly amenable to the PIC approach \cite{juno2018discontinuous,von2014vlasiator}.

It is well known that the noise in the PIC simulations may significantly undermine the physical results of the simulations. For example, the PIC simulations of electron temperature gradient modes\cite{lin2005role,lin2005Electron} yielded a level of turbulent heat transport that deviated greatly from results of gyrokinetic Vlasov simulations of Refs.~\onlinecite{dorland2000electron,jenko2000electron,jenko2002prediction}. The origin of this discrepancy is investigated in  Ref.~\onlinecite{nevins2005discrete}. It is shown that the discrete particle noise effects in the PIC simulations of Refs.~\onlinecite{lin2005role,lin2005Electron} undermine the dynamics of the instability, strongly modifying the predictions for the heat transport levels \cite{nevins2005discrete,holod2007statistical}. The role of the PIC noise has also been discussed in the study of electrodynamic filamentation instability \cite{palodhi2019counterstreaming}. It is shown that the noise in PIC simulation affects the mechanism of the instability and results in an incorrect instability threshold. In another study, it is shown that the noise of the PIC simulations can lead to significant artificial heating of plasma in the presence of the Monte Carlo collision operator \cite{turner2006kinetic}. In Refs.~\onlinecite{nevins2006characterizing,juno2020noise,holod2007statistical}, it is emphasized that the role of the discrete particle noise in PIC simulations has to be carefully analyzed and evaluated for each physical situation. For this purpose, several approaches have been proposed in the literature\cite{riva2017methodology,holod2007statistical,kesting2015propagation,turner2016verification,turner2013simulation}. One approach is to benchmark the physical results with different simulation methods in order to build confidence in simulation results and determine the roots of discrepancies and possible numerical artifacts. Benchmarking has been successfully used as a verification tool for numerical codes in several publications~\cite{turner2013simulation,villafana20212d,charoy20192d}. 
One feature of benchmarking is that it tests the entire simulation code as opposed to individual units, and it can also be used on the specific problems of interest rather than synthetic test cases~\cite{turner2013simulation,turner2016verification}. Benchmarking with  different  numerical methods, such as PIC and Vlasov methods, provides additional confidence in the reliability of the simulations as well as highlighting the causes of possible discrepancies.   

In this study, we use several PIC and Vlasov codes to investigate the impact of noise in  PIC simulations of the Buneman instability. We show that the linear growth rates of the instability are significantly affected by the noise inherent to the PIC simulations. We identify  the trapping of electrons (a nonlinear effect) in the early noise-driven potential as a source of the inconsistencies with the linear theory. This relationship is confirmed by continuum (Vlasov) simulations for the same parameters and initial states (and respectively the same level of noise) as in the corresponding PIC simulations. It is also supported by analytical calculations that show a high sensitivity of the linear growth rates in this problem  to small distortions of the Maxwellian velocity distribution function. Therefore, we propose that early trapping of electrons induces a small plateau in the velocity distribution function, leading to the much higher linear growth rates observed in PIC simulations. 

The similarities and differences between the PIC and Vlasov simulations are presented through a number of simulations. We begin with PIC simulations of the cold-plasma limit, when $v_0=6v_{te}$ is relatively large. In this case, the simulated growth rates are shown to be consistent with the theoretical ones. 
A set of simulations is then presented for a relatively low value of the streaming velocity, $v_0=2v_{te}$, where $v_{te}=\sqrt{T_e/m_e}$ is the thermal velocity of the electrons, $T_e$ is the initial temperature of the electrons, and $m_e$ is the electron mass. In each simulation, we measure the linear growth rates of several modes and compare them with the results of the linear theory. Some Vlasov simulations are started with an extremely small perturbation that is required by this method to excite the instability. We refer to these simulations as ``low-noise" Vlasov simulations (VL1 and VL2 in \Cref{table:simlist}). The growth rates measured by the low-noise Vlasov simulations are shown to be consistent with the linear theory. On the other hand, some PIC simulations are started with macroparticles randomly distributed in phase space. We refer to these simulations as ``random-start" PIC simulations (PIC1, PIC2, PIC3, PIC4, and PIC 5 in \Cref{table:simlist}). The growth rates measured using random-start PIC simulations deviate significantly (up to a factor of 3) from the linear theory. This discrepancy in linear growth rates persists in the random-start PIC simulations using up to $10^5$ macroparticles per cell. In addition, we show that starting a Vlasov simulation with the same level of initial noise as the PIC simulations (VL3 in \Cref{table:simlist})  leads to a similar discrepancy between the simulated and the theoretical growth rates. 

Reflecting its statistical origin, the noise in PIC simulations scales as $1/\sqrt{N_p}$, where $N_p$ is the number of macroparticles in each grid cell.  The initial noise is a result of the random distribution of the macroparticles in phase space before the first time step. To reduce the adverse effects of the initial noise, a ``quiet-start" initialization has been proposed\cite{dawson1983particle,byers1970perpendicularly}. In contrast to the random-start method, in the quiet-start method, the macroparticles are distributed regularly or semi-regularly with appropriate weights in phase space. Accordingly, the initial noise level is made much smaller.
We show that using the quiet-start method does not completely solve the noise problem in PIC simulations. Another outcome of the current study is to show how the quiet-start method should be used to improve the accuracy of the observed linear growth rates in PIC simulations. We first show that although the quiet-start can improve the results by reducing initial noise, the growth of modes is still subject to statistical noise, making an accurate measurement of the linear growth rates difficult. However, initially perturbing the mode with the maximum growth rate helps to achieve better consistency with the linear theory. 

The outline of the next sections is as follows. In \cref{sec:linear_theory}, we review the linear theory of our problem and introduce the general setup for the simulations. In \cref{sec:large_v0}, we show some results for a large $v_0$ ($v_0=6v_{te}$) value that show good agreement between the theoretical linear growth rates and the growth rates measured from the theory. In \cref{sec:low_v0}, we show various simulations with the PIC and Vlasov methods. As a result, we show how the initial noise of random-start PIC simulations adversely influences with the linear growth and undermines the accuracy of growth rate measurements, a problem that does not appear in the low-noise Vlasov simulations. In \cref{sec:small_flattening}, we show that a small flattening in the distribution function can increase the observed linear growth rates by several factors. This provides a hypothesis as to source of the problem in random-start PIC simulations. In \cref{sec:quite_start}, we show that, although it does lead to some improvements, the quiet-start method is unlikely to completely solve the problem of the noise in PIC simulation. In \cref{sec:conclusion}, we summarize the conclusions of this study. 

\section{The Buneman instability and the problem setup}\label{sec:linear_theory}

The Buneman-type instabilities are driven by the relative drift $v_0$ of electrons with respect to ions in an unmagnetized plasma. The instabilities can be categorized according to the magnitude of $v_0$\cite{galeev1984current}. The Buneman instability regime occurs for  $v_0>v_{te}$. On the other hand, the ion-sound instability occurs for  $v_{ti}<v_0<v_{te}$, where $v_{ti}=\sqrt{T_i/m_i}$, $T_i$ is the initial temperature of the electrons, and $m_i$ is the ion mass. Streaming Buneman-type instabilities are important in many topical problems of plasma physics. For example, they are considered as candidates for explaining the turbulence and anomalous resistivity in solar plasmas\cite{ChePoP2017,HellingerGRL2004,CheHH_MPLA2016} and hollow cathode plasmas in Hall thrusters \cite{HaraPSST2019} as well as sources of nonlinear effects in  ion-beam fusion applications\cite{StartsevPoP2006}. 

The Buneman instability has been broadly investigated through  numerical simulation \cite{lampe1974two, niknam2011simulation,rajawat2017particle,tavassoli2021backward}.
Most of the numerical simulations focus on the nonlinear regimes of the instability, assuming that the linear regime is well understood via  analytical  dispersion relations. However, the comparison of the linear regime in numerical simulations with the  linear theory provides a valuable test for the  simulation methods, revealing the validity range of the linear approximation for a particular approach. In this study,  we focus on the verification of the linear regime of the Buneman instability in PIC and Vlasov simulations. 

The considered equations in the setup of our problem are
\begin{align}
\pdv{f_{i,e}}{t}+v\pdv{f_{i,e}}{x}+ \frac{qE_x}{m_{i,e}}\pdv{f_{i,e}}{v}&=0,\nonumber\\
\pdv{E_x}{x}&=e(n_i-n_e),
\end{align}
 where $f_{i,e}$ is the distribution function for ions and electrons, respectively, $E_x$ is the electric field, $n_{i,e}=\int f_{i,e}\,\dd v$ are the ion and electron densities, and $q$ is the charge, which is $+e$ for the ions and $-e$ for the electrons. The ions are taken to be Hydrogen with mass $m_i=1$ amu. The initial temperature for both ions and electrons is $T_0=0.2$ eV; the initial plasma density is $n_0=10^{17}\;\text{m}^{-3}$.
 The initial conditions are
\begin{align}
f_i(x,v,0)&=\frac{n_0}{\sqrt{2\pi}\,v_{ti}}\exp(-\frac{v^2}{2v_{ti}^2}),\\
f_e(x,v,0)&=\frac{n_0(1+\epsilon \cos{(k_0x)})}{\sqrt{2\pi}\,v_{te}}\exp{-\frac{(v-v_0)^2}{2v_{te}^2}}. \label{electron_init_VDF}
\end{align}
  The quantities $\epsilon$, and $k_0$ parameterize a small initial perturbation. In the low-noise Vlasov and quiet-start PIC  simulations, these parameters are required to excite the instability. In the low-noise Vlasov simulations, we take $\epsilon=10^{-5}$, while in the quiet-start PIC simulations we take $\epsilon=10^{-8}$. In the random-start PIC simulations, there is no need for this perturbation, and we take $\epsilon=0$. In all the simulations reported, we use  periodic boundary conditions in a system of length 6 mm discretized with a grid of 2048 points. This length is large enough to allow excitation of several modes with mode numbers $m\equiv kL/2\pi$. The time step used in the simulations is $\Delta t=2.39\times 10^{-4}$ ns. All the time-dependent data are collected at intervals of 500 $\Delta t$. The relative drift between ions and electrons ($v_0$) drives the instability in several modes identified by the linear dispersion relation
\begin{gather}
    1-\frac{\omega_{pi}^2}{2k^2v_{ti}^2}Z^{\prime}\qty(\frac{\omega}{\sqrt{2}\abs{k}v_{ti}})-\frac{\omega_{pe}^2}{2k^2v_{te}^2}Z^{\prime}\qty(\frac{\omega-kv_{0}}{\sqrt{2}\abs{k}v_{te}})
    =0,
    \label{dispersion}
\end{gather}
where $\omega\equiv \omega_R+i\gamma$ with $\gamma$ the linear growth rate, $\omega_R$ the frequency, $k$ is the wave vector, $\omega_{pi}$ is the ion plasma frequency, $\omega_{pe}$ is the electron plasma frequency, and $Z$ is the plasma dispersion function.

\section{Linear growth rates from PIC simulations for large drift velocity,  $v_0=6v_{te}$.}\label{sec:large_v0}

\begin{figure}[htbp]
\centering
\captionsetup[subfigure]{labelformat=empty}
\subcaptionbox{\label{E_k_growth_6vth_10000}}{\includegraphics[width=.46\linewidth]{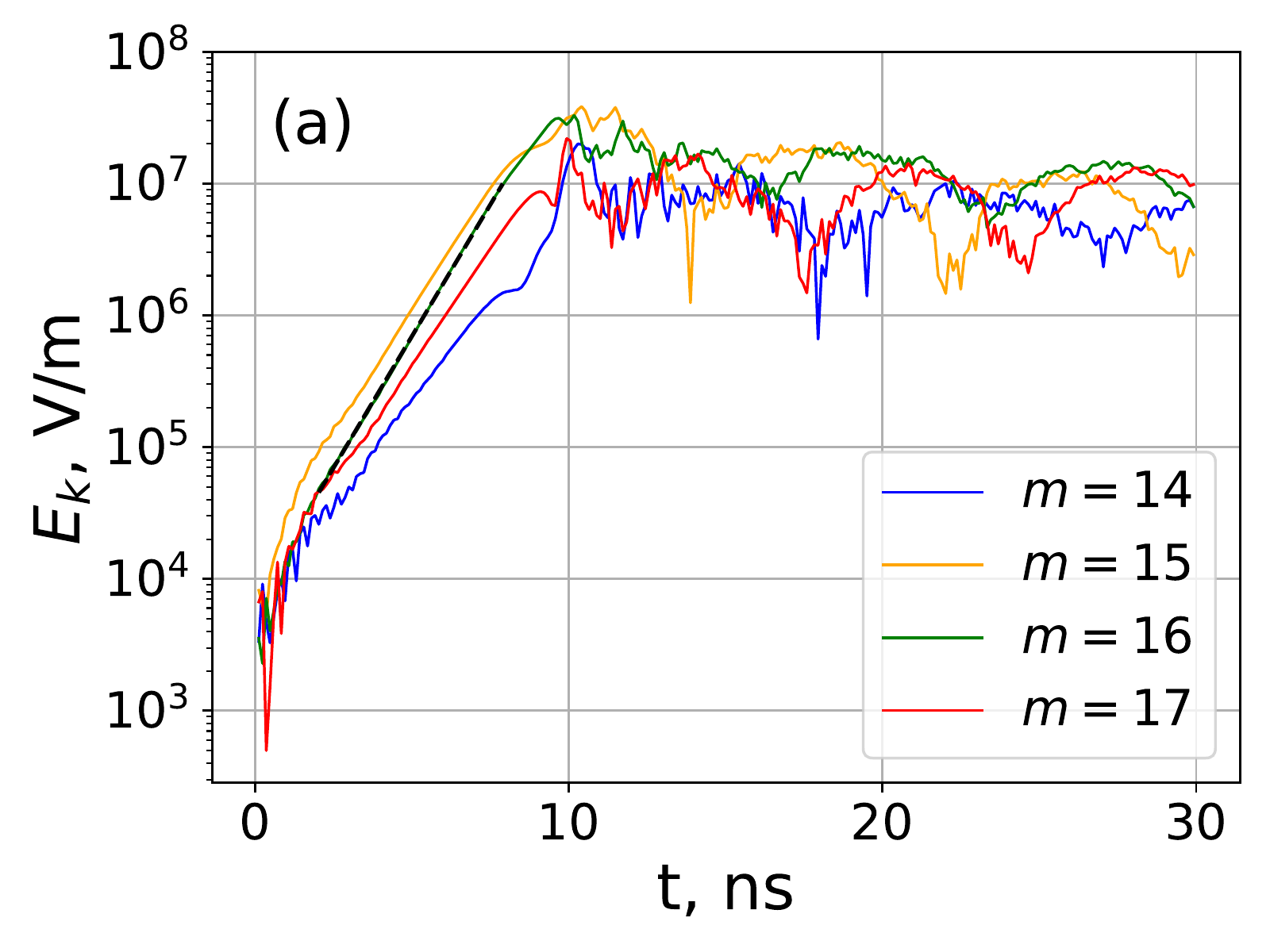}}
\subcaptionbox{\label{es_10000_6vte}}{\includegraphics[width=.46\linewidth]{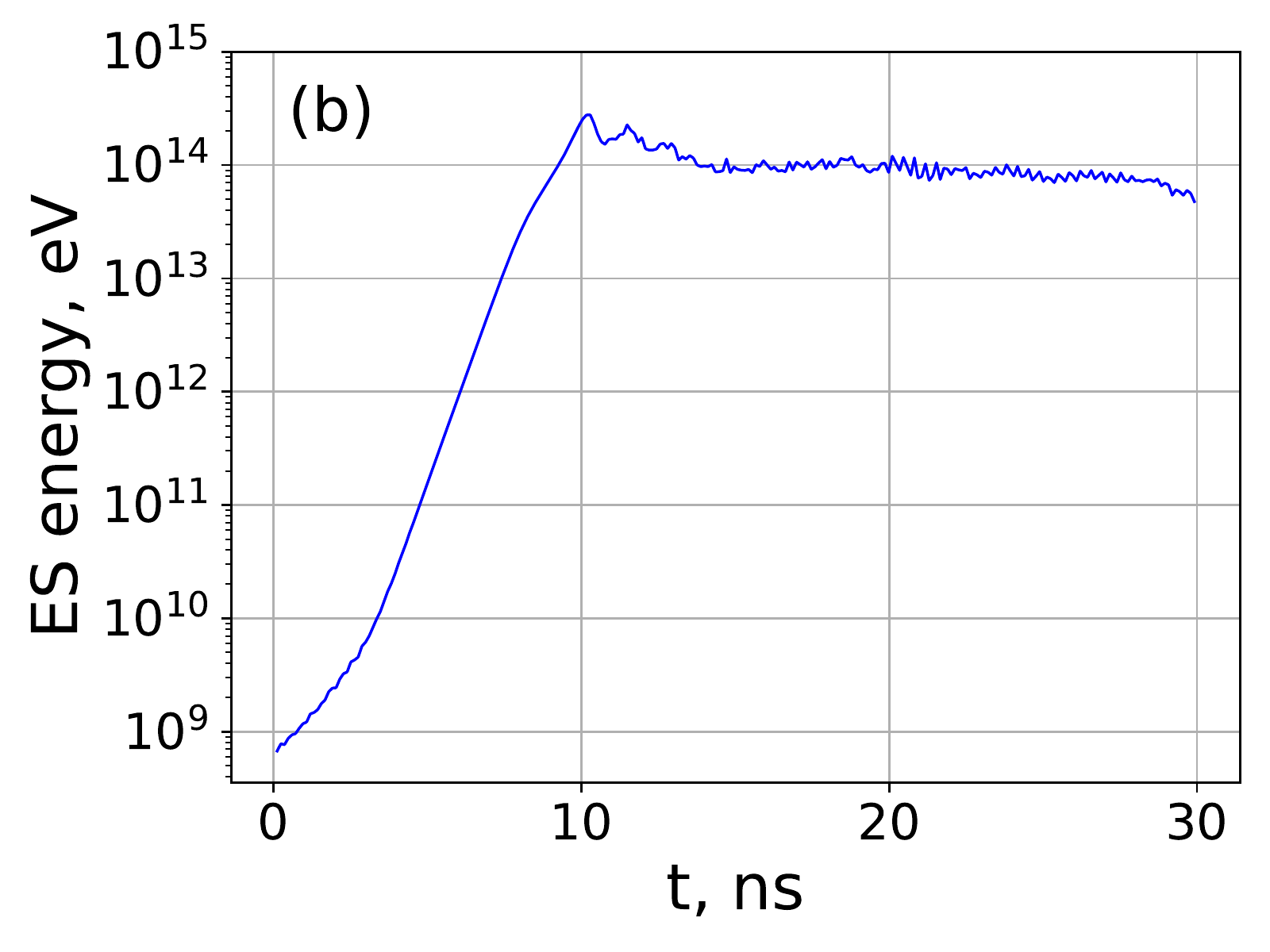}}
\caption{a) The evolution of individual modes of the  electric field. The dashed black line shows the fitted line on the $m=16$ mode. b) The evolution of the electrostatic energy. Both figures are from VSim PIC simulations for the case $v_0=6v_{te}$ (see below).}
\end{figure}

By choosing the drift velocity of $v_0=6v_{te}$, we approach the cold-plasma limit of the Buneman instability.  We perform PIC simulations in this limit with $10^4$ macroparticles per cell. \Cref{E_k_growth_6vth_10000} shows the growth of some select modes. These modes are chosen for the linear growth analysis of the case $v_0=6v_{te}$ and include the maximum growth rate mode $m=16$. We can see a distinct linear growth region in the early evolution of the modes. By fitting a line to this region, we calculate the growth rate of each mode. In \Cref{table:growth_numbers_6vte}, the calculated growth rates are shown to be quite consistent with the results from the linear theory. The standard error (SE) associated with the measurement of the growth rates is also reported in this table. The SEs of the fits are quite small, showing that the growth of the chosen modes is quite linear and not oscillatory. Our other investigations (not reported here) show that even for as few as $10^3$ macroparticles per cell, PIC simulations with $v_0=6v_{te}$ produce accurate linear growth rates.

The mean of the derivative of the spectral growth over the same time period is an equivalent measure of the growth rate. Variations about the mean provide a measure of the power of the noise present in the growth region. The square of the growth rate over the square of the power of noise was calculated as the signal-to-noise ratio (SNR) of the growth rate \cite{Lynn1989}. The average SNR of the chosen modes is 21.25 dB, which is much greater than 1. This indicates that the power of noise carried in this case is quite low in the linear growth region. In all the simulations reported in this study, we see that the value of the SNR does not vary much among the chosen modes. Therefore, we only report the SNR averaged over the four chosen modes of each simulation.

\begin{table}[htbp]
\begin{tabular}[b]{|c|c|c|c|}
\hline
$m$  & \begin{tabular}[c]{@{}c@{}}$\gamma$ (Theory)\\ $\times 10^8$ s$^{-1}$\end{tabular} & \begin{tabular}[c]{@{}c@{}}$\gamma$ (Simulation)\\ $\times 10^8$ s$^{-1}$\end{tabular}&\begin{tabular}[c]{@{}c@{}}SE (Simulation)\\ \%\end{tabular} \\ \hline
14 & 7.08                                                                            & 7.09 &  1.03                \\ \hline
15 & 8.67                                                                            & 8.60 &  0.30                                                                   \\ \hline
16 & 9.46                                                                            & 9.20  &  0.24                                                                    \\ \hline
17 & 8.13                                                                           & 8.12 &  0.10                                                                     \\ \hline
\hline
Average & 8.34 & 8.25 & 0.42 \\ 
\hline
\end{tabular}
\caption{Comparison of theoretical growth rates with growth rates observed in VSim PIC simulations with $v_0=6v_{te}$.\label{table:growth_numbers_6vte}}
\end{table}

\section{Linear growth rates from Vlasov and random-start PIC simulations for low drift velocity,  $v_0=2v_{te}$.}
\label{sec:low_v0}

In this section, we report on linear growth rates from several PIC and Vlasov simulations for the case of $v_0=2v_{te}$. As we show, this regime of relatively low drift velocity can be problematic for the PIC simulations. Therefore, we investigate this regime more extensively by performing several PIC and Vlasov simulations. Due to the large number of these simulations, we assign a specific name to each one in this regime. These simulations are listed and described in \Cref{table:simlist}.

\begin{table}[htbp]
\begin{tabular}{|c|c|c|c|}
\hline
Simulation & Numerical code         & Initial condition                        & macroparticles per cell \\ \hline
VL1        & Semi-Lagrangian Vlasov  & $m=1$ perturbed                         & ---                       \\ \hline
VL2        & BOUT++                 & $m=1$ perturbed                         & ---                       \\ \hline
PIC1       & EDIPIC                 & Random start, no perturbation            & $10^4$                   \\ \hline
PIC2       & VSim                   & Random start, no perturbation            & $10^4$                   \\ \hline
PIC3       & XES1                   & Random start, no perturbation            & $10^4$                   \\ \hline
PIC4       & EDIPIC                 & Random start, no perturbation            & $10^5$                  \\ \hline
PIC5       & VSim                   & Random start, no perturbation            & $10^5$                  \\ \hline
VL3        & Semi-Lagrangian Vlasov & Identical to PIC2         & ---                       \\ \hline
PIC6       & EDIPIC                 & Quiet-start, $m=44$ perturbed           & $10^4$                   \\ \hline
PIC7       & VSim                   & Quiet-start,  $m=44$ perturbed           & $10^4$                   \\ \hline
PIC8       & XES1                   & Quiet-start,  $m=44$ perturbed           & $10^4$                   \\ \hline
PIC9      & VSim                   & Quiet-start, $m=\{31,37,44,51\}$ perturbed & $10^4$                   \\ \hline
PIC10       & EDIPIC                   & Quiet-start, no perturbation             & $10^4$                   \\ \hline
PIC11      & EDIPIC                   & Quiet-start, $m=1$ perturbed            & $10^4$                   \\ \hline
PIC12      & EDIPIC                   & Quiet-start,  $m=31$ perturbed           & $10^4$                   \\ \hline
\end{tabular}
\caption{The list of simulations with $v_0=2v_{te}$.}
\label{table:simlist}
\end{table}

The first simulation (VL1) is performed by a locally developed semi-Lagrangian code. The semi-Lagrangian Vlasov scheme is a well-known and tested scheme for solving the Vlasov--Poisson equations\cite{cheng1976integration,gagne1977splitting}. In this scheme, the Vlasov equation is split into a convection equation and a force equation. Each of these equations is then solved by the method of characteristics using cubic spline interpolation. The Poisson equation is solved by a spectral method, the FFT. The second Vlasov simulation (VL2) is done with the BOUT++ framework. BOUT++ is a modular platform for 3D simulations of an arbitrary number of fluid equations in curvilinear coordinates using finite-difference methods \cite{dudson2009bout++, dudson2015bout++}. Time integration of partial differential equations (PDEs) in BOUT++ is based on the method of lines. The time stepping is performed with the CVODE ODE solver from the SUNDIALS package~\cite{hindmarsh2005sundials} using variable-order, variable-step multistep methods and is suitable for stiff and nonstiff problems. Spatial derivatives are treated with the third-order weighted essentially non-oscillatory (WENO) scheme for upwind terms and a fourth-order central-difference scheme for other first-order derivatives. In the Vlasov simulations, the velocity boundary conditions are open, and the velocity grid consists of 2001 points. This leads to a velocity resolution of $0.027\;c_s$ and $0.53\;c_s$ for the ions and electrons, respectively, where $c_s=\sqrt{T_0/m_i}$ is the ion sound velocity. We start the low-noise Vlasov simulations (VL1 and VL2) with an extremely small initial perturbation ($\epsilon=10^{-5}$). 

\begin{figure}[htbp]
\centering
\captionsetup[subfigure]{labelformat=empty}
\subcaptionbox{\label{t_ES}}{\includegraphics[width=.46\linewidth]{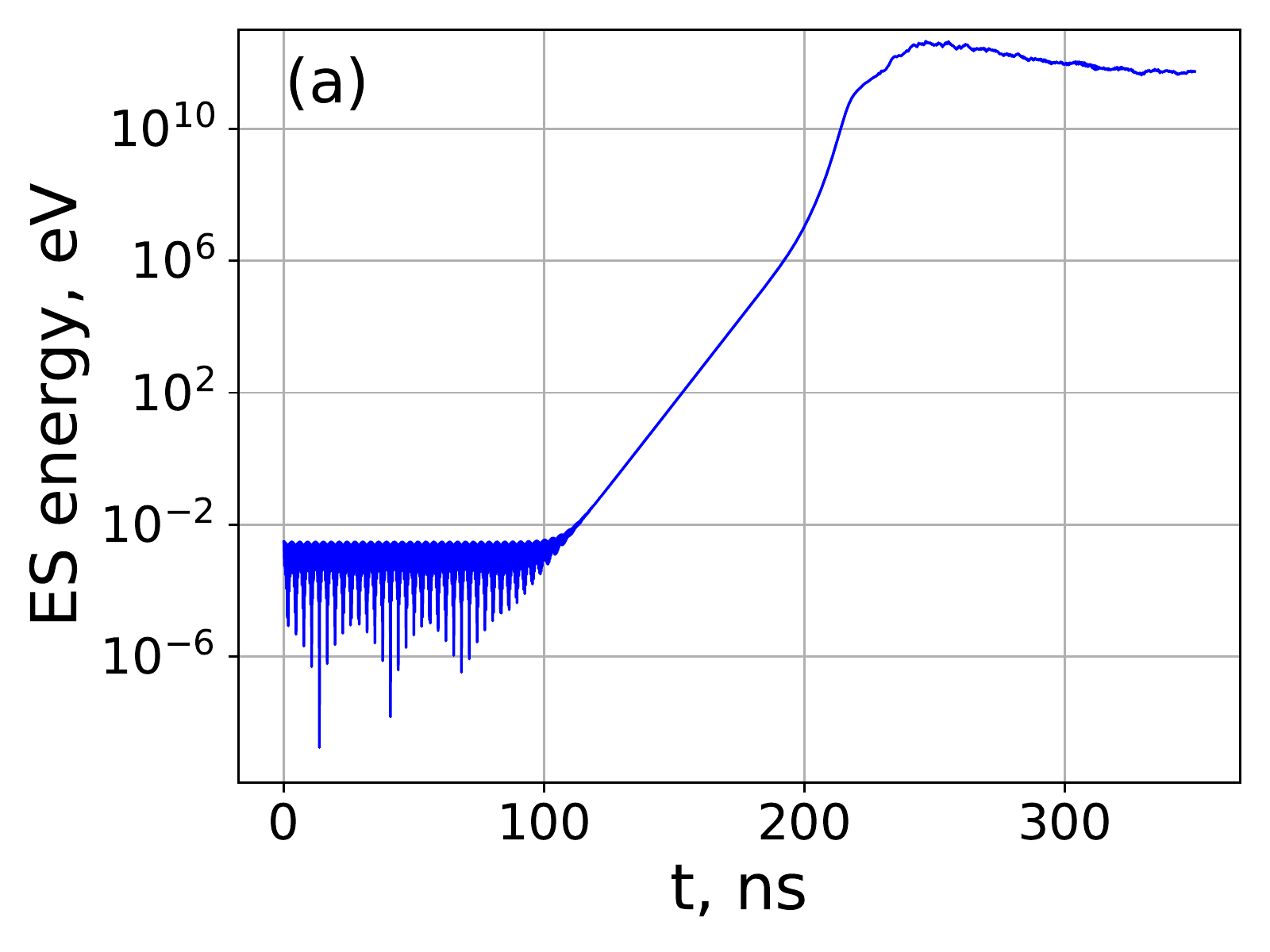}}
\subcaptionbox{\label{es_energydata}}{\includegraphics[width=.46\linewidth]{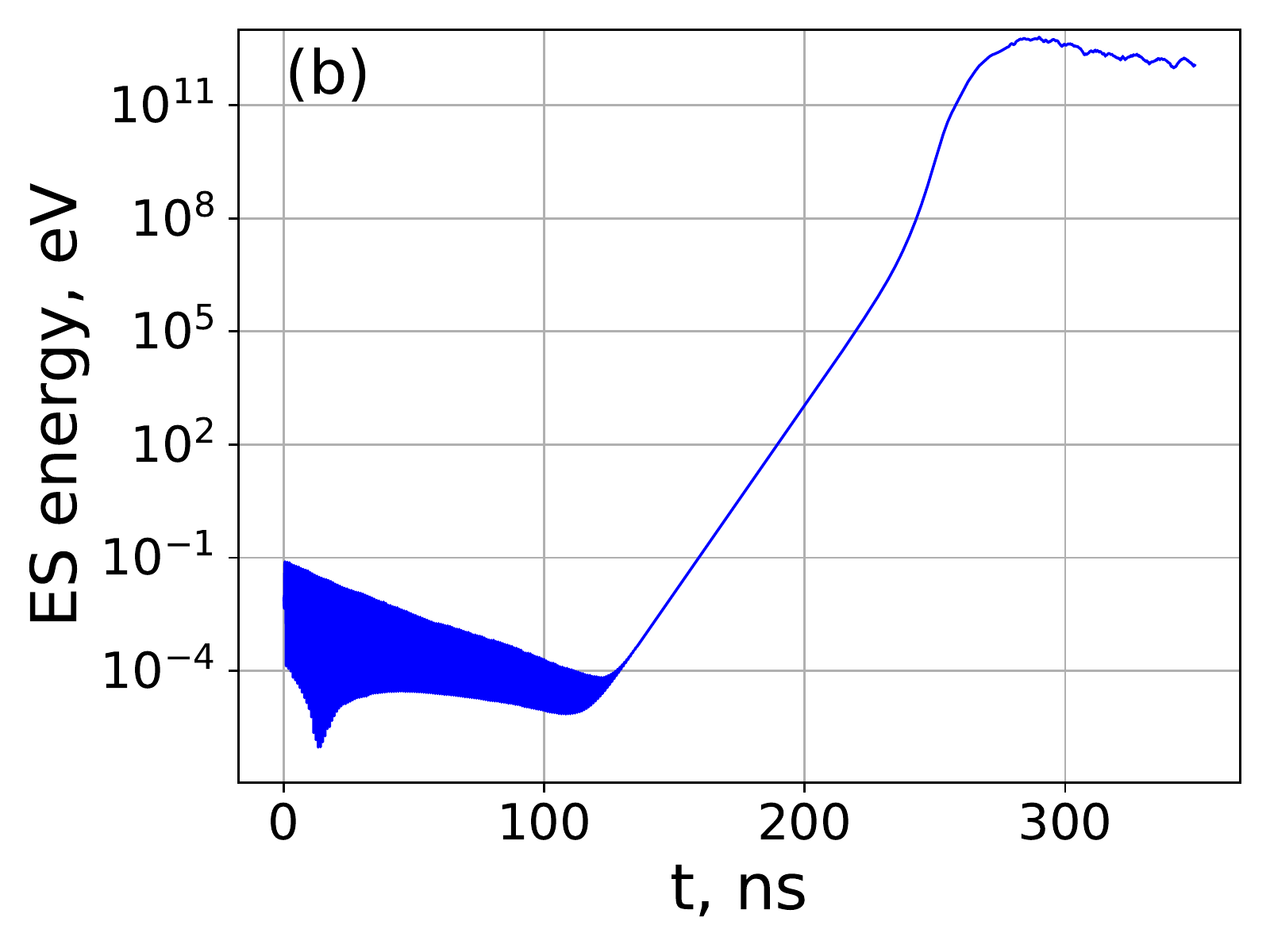}}
\caption{The evolution of the electrostatic energy in the low-noise Vlasov simulations (VL1 and VL2). a)  Semi-Lagrangian (VL1) and b) BOUT++ (VL2).}
\label{low_noise_ES}
\end{figure}

\begin{figure}[htbp]
\centering
\captionsetup[subfigure]{labelformat=empty}
\subcaptionbox{\label{t_Ek_correct}}{\includegraphics[width=.46\linewidth]{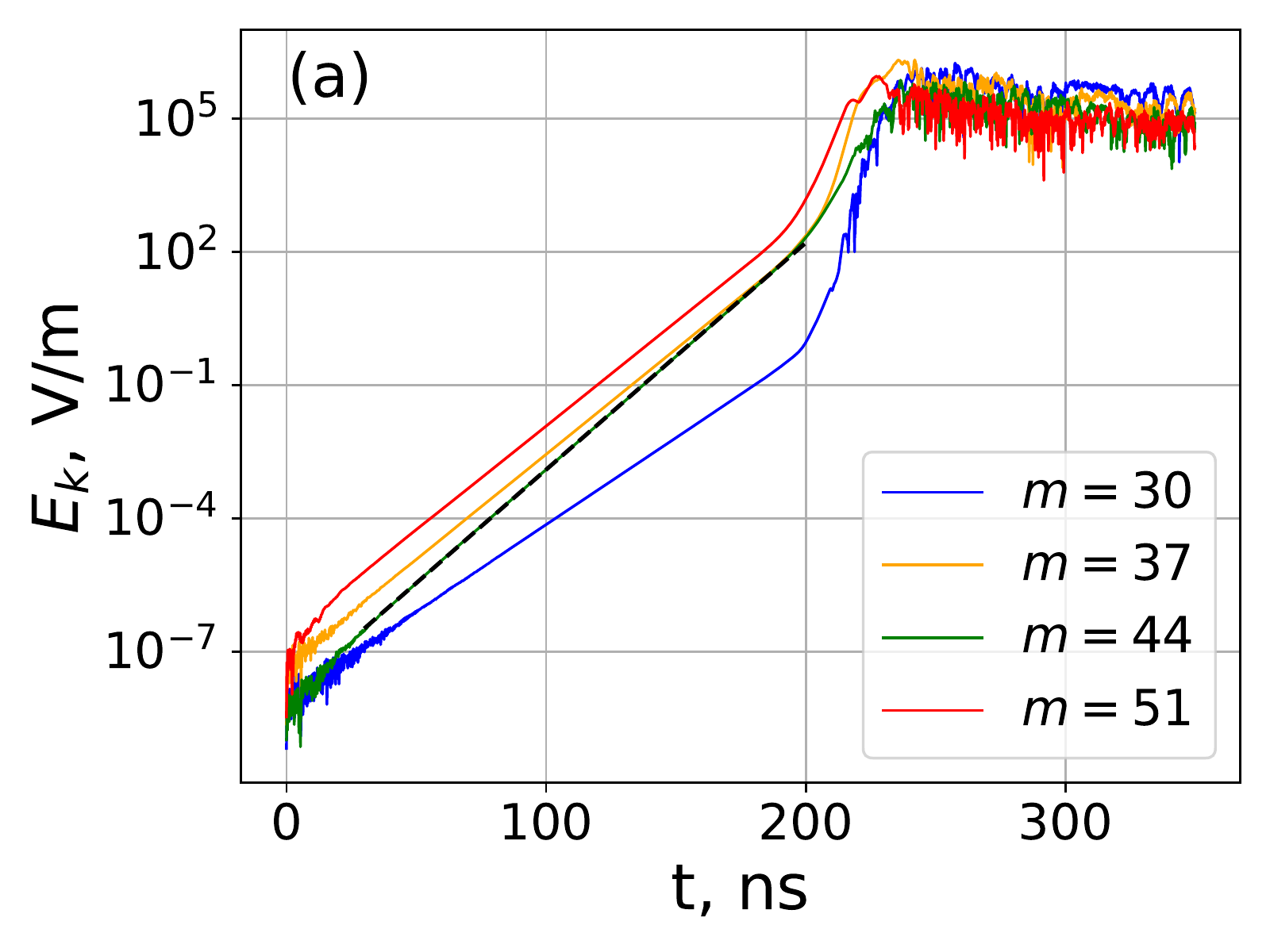}}
\subcaptionbox{\label{e_modes}}{\includegraphics[width=.46\linewidth]{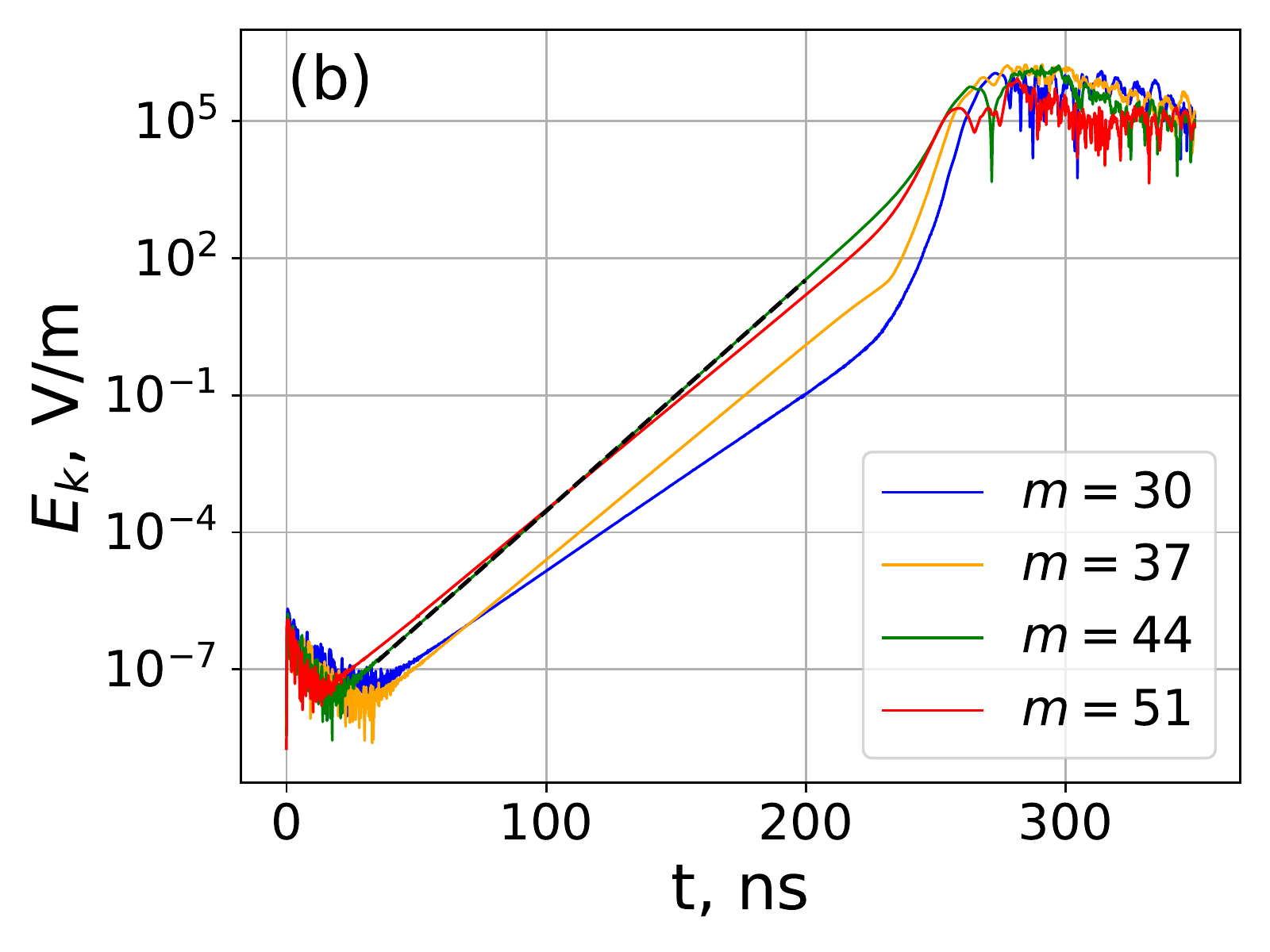}}
\caption{a) The evolution of individual modes of the electric field in the low-noise Vlasov simulations (VL1 and VL2). a)  semi-Lagrangian (VL1) and b) BOUT++ (VL2). The dashed black line shows the fitted line for the $m=44$ mode.}
\label{low_noise_Ek}
\end{figure}

\Cref{low_noise_ES}~shows the evolution of electrostatic (ES) energy in the low-noise Vlasov simulations (VL1 and VL2), and from here, the linear growth and the nonlinear saturation can be seen. The ES energy in the VL2 simulation starts growing from a larger value than the VL1 simulation. This difference is likely due to the Poisson solver used in the BOUT++ code that introduces some initial noise that is not present in the semi-Lagrangian code. The ES energy in VL2 simulation, however, damps to a value close to the starting energy in VL1 after about 100 ns. This damping leads to some phase difference between the two simulations, so that after 350 ns, the ES energy is higher in the VL2 simulation. The linear growth regimes, which come after about 100 ns in VL1 and 125 ns in VL2, are highly similar in both simulations.
For the calculation of linear growth rates, we have chosen four individual modes of the electric field. These modes are $m=\{30,37,44,51\}$ in all simulations (PIC and Vlasov) for the case $v_0=2v_{te}$. According to  the linear theory, mode $m=44$ has the maximum linear growth rate in our setup. In \Cref{low_noise_Ek}, the linear growth region is clearly seen for each mode. \Cref{table:growth_numbers_2vte_VL} shows the values of the linear growth rates calculated from the low-noise Vlasov simulations and the linear theory are quite consistent with each other. The low SEs reported in \Cref{table:growth_numbers_2vte_VL} reflect the fact that the growth is essentially linear. The average SNR of the chosen modes in the linear growth region are 49.13 dB in the VL1 simulation and  17.75 dB in the VL2 simulation. Because the SNR in both simulations is much greater than unity, we can say the power of noise carried in the growth region is quite small.

\begin{table}[htbp]
\begin{tabular}[b]{|c|c|c|c|c|c|}
\hline
$m$  & \begin{tabular}[c]{@{}c@{}}$\gamma$ (Theory)\\ $\times 10^8$ s$^{-1}$\end{tabular} & \begin{tabular}[c]{@{}c@{}}$\gamma$ (VL1)\\ $\times 10^8$  s$^{-1}$\end{tabular}&\begin{tabular}[c]{@{}c@{}}SE (VL1)\\ \%\end{tabular} &\begin{tabular}[c]{@{}c@{}}$\gamma$ (VL2)\\ $\times 10^8$  s$^{-1}$\end{tabular}&\begin{tabular}[c]{@{}c@{}}SE (VL2)\\ \%\end{tabular} \\ \hline
30 & 0.90                                                                            & 0.9 &  0.029   & 0.90   &   0.014                                                                    \\ \hline
37 & 1.08                                                                            & 1.09 &   0.004   & 1.08  &  0.006                                                                     \\ \hline
44 & 1.17                                                                            & 1.17 &  0.017   & 1.16  &  0.009                                                                   \\ \hline
51 & 1.07                                                                            & 1.08 &  0.004   & 1.08 &   0.006                                                                     \\ \hline
\hline
Average & 1.06 & 1.06 & 0.014 & 1.06 & 0.009 \\
\hline
\end{tabular}
\caption{Comparison of the theoretical growth rates with the growth rates observed in the VL1 and  VL2 simulations.\label{table:growth_numbers_2vte_VL}}
\end{table}

\begin{figure}[htbp]
\centering
\captionsetup[subfigure]{labelformat=empty}
\subcaptionbox{\label{EdiPic_10kmp_2vte_EsPe}}{\includegraphics[width=.49\linewidth]{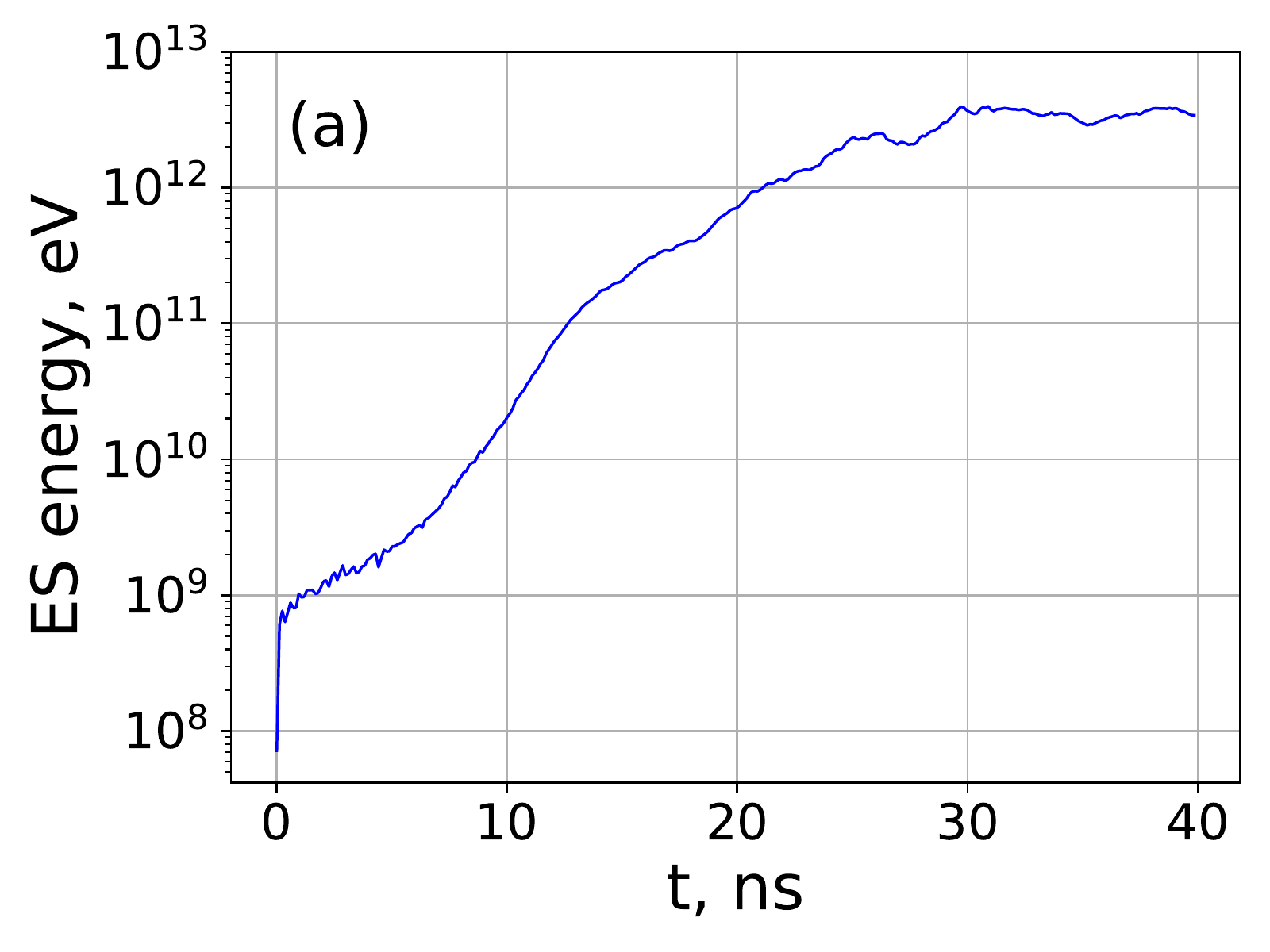}}
\subcaptionbox{\label{es_10000}}{\includegraphics[width=.49\linewidth]{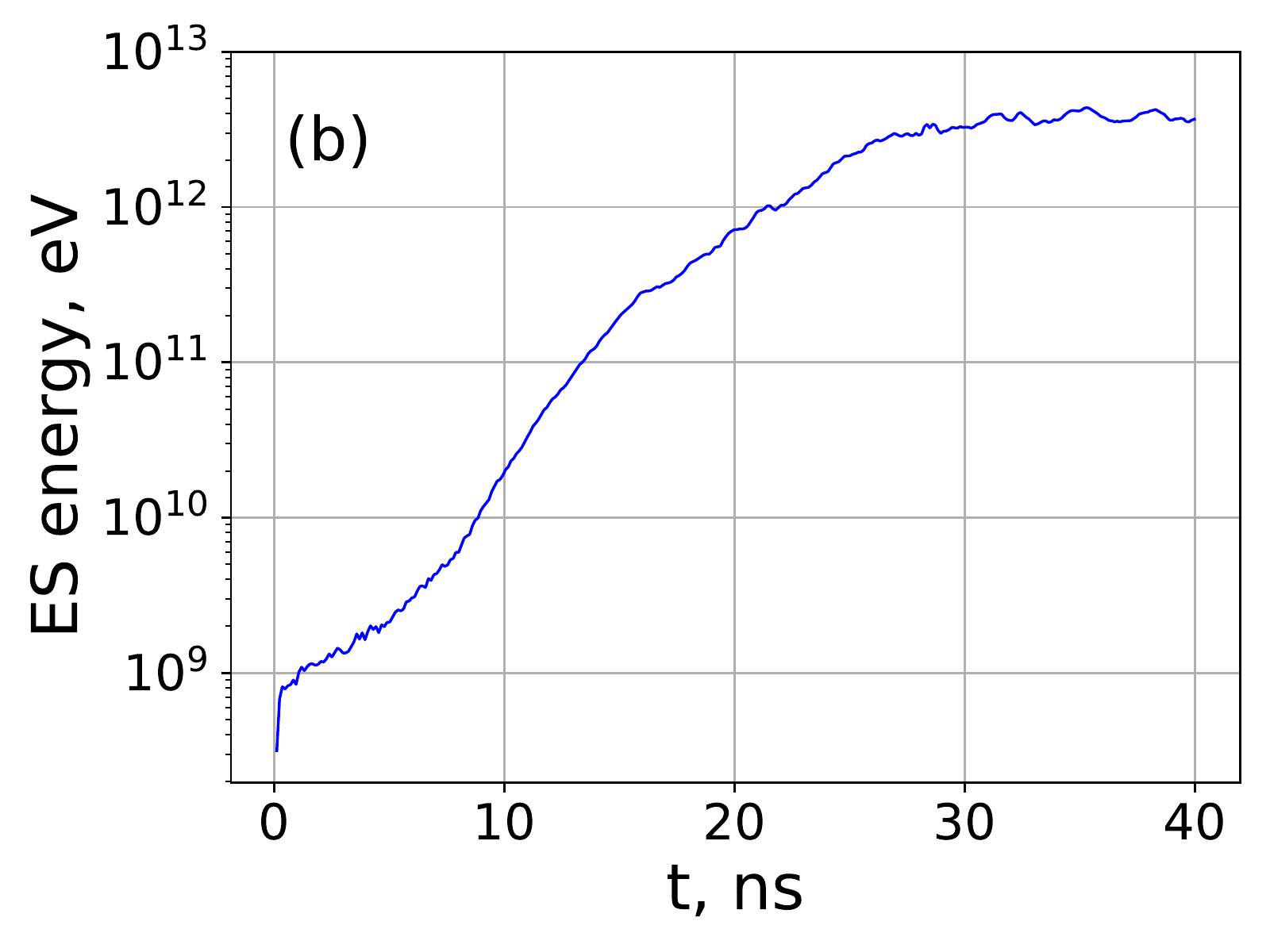}}

\subcaptionbox{\label{X-wqs-es-10kppc}}{\includegraphics[width=.49\linewidth]{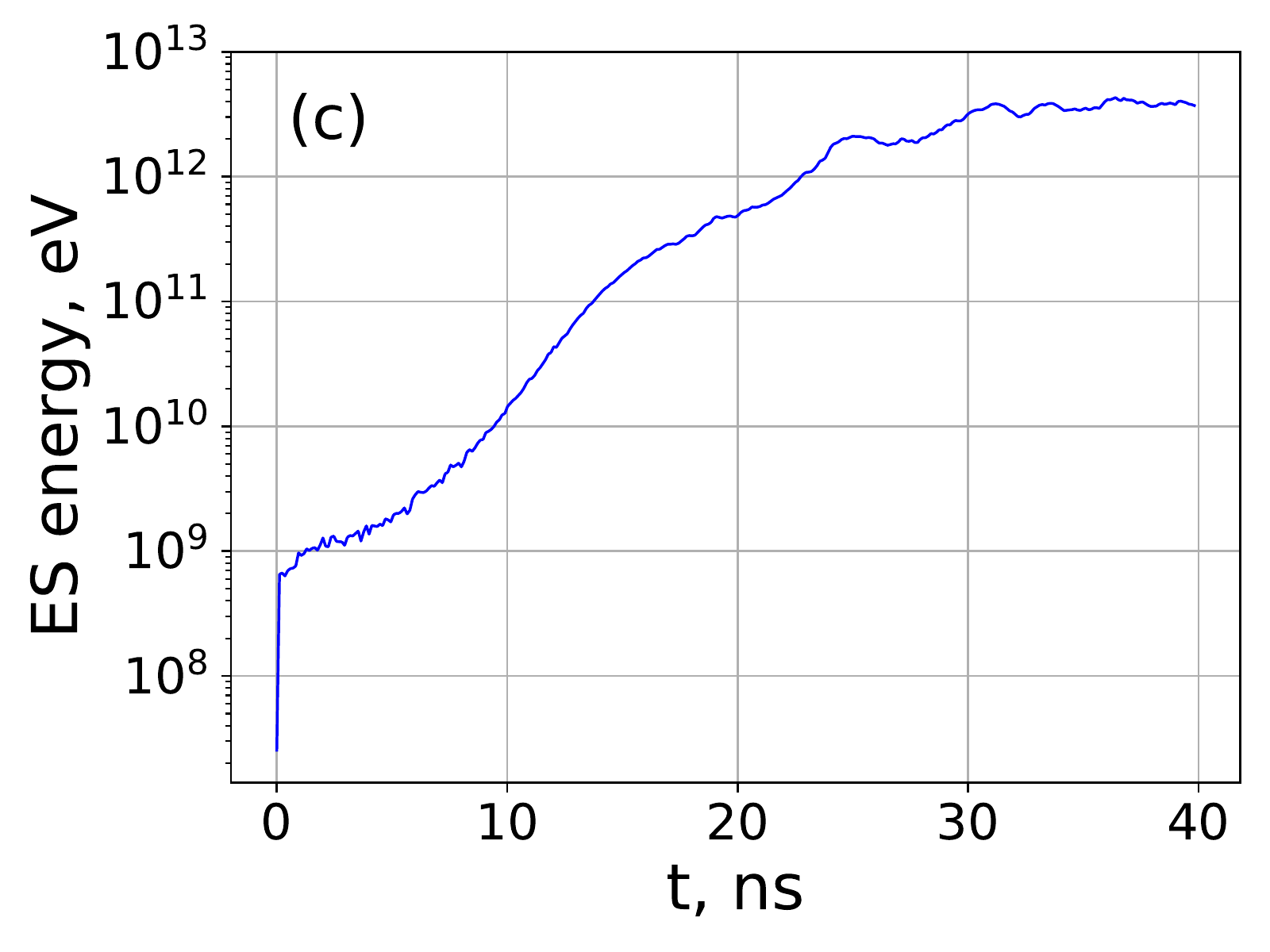}}
\caption{The evolution of the electrostatic energy in random-start PIC using $10^4$ macroparticles per cell from a) EDIPIC (PIC1) b) VSim (PIC2) c) XES1 (PIC3) simulation.}
\label{random_start_1e4_ES}
\end{figure}

\begin{figure}[htbp]
\centering
\captionsetup[subfigure]{labelformat=empty}
\subcaptionbox{\label{EdiPic_10kmp_2vte_Ex_Growth}}{\includegraphics[width=.49\linewidth]{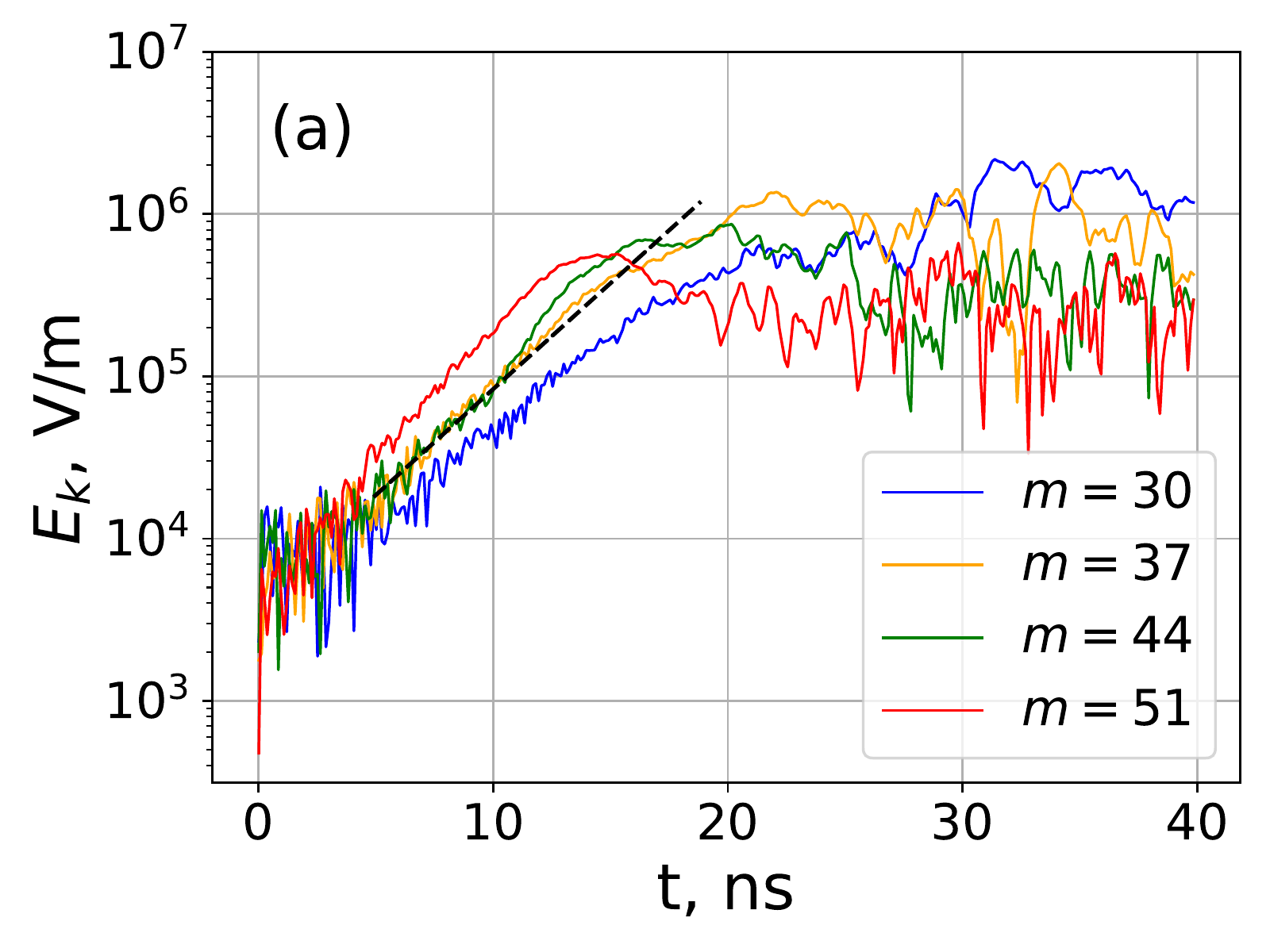}}
\subcaptionbox{\label{E_k_growth_10000}}{\includegraphics[width=.49\linewidth]{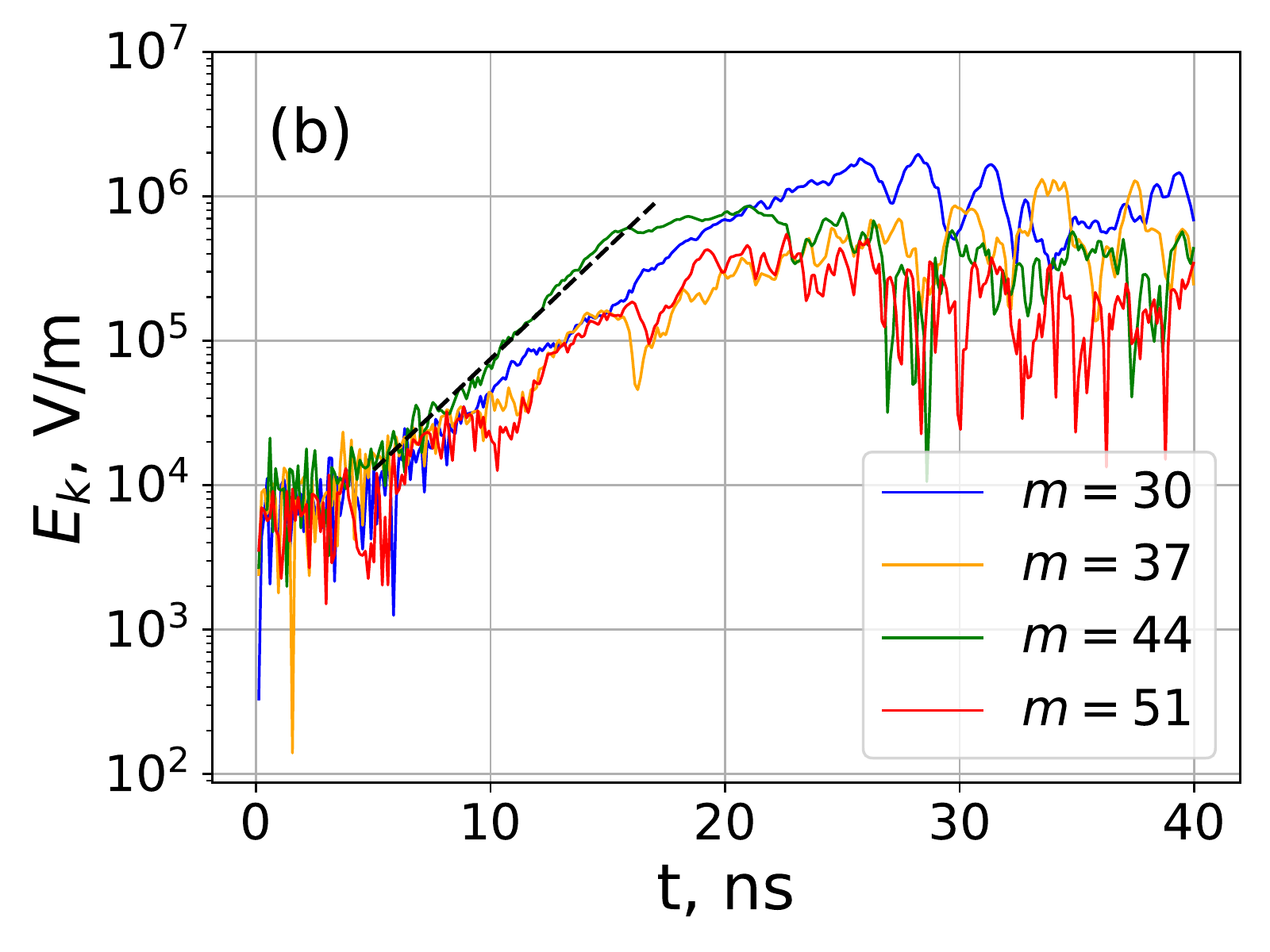}}

\subcaptionbox{\label{X-wqs-10kppc}}{\includegraphics[width=.49\linewidth]{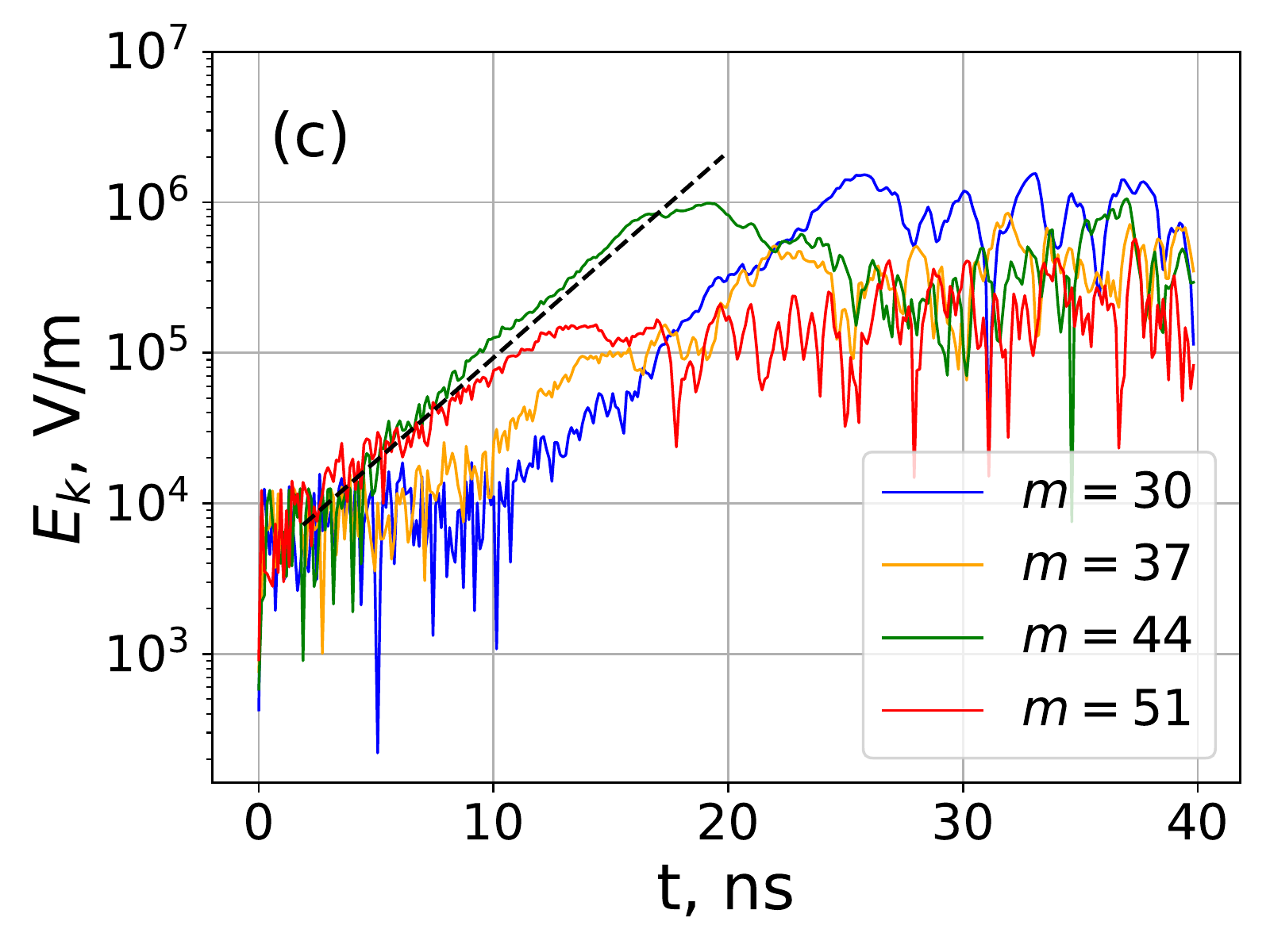}}
\caption{The evolution of individual modes of the electric field in random-start PIC simulations using $10^4$ macroparticles per cell from a) EDIPIC (PIC1), b) VSim (PIC2), and c) XES1 (PIC3) simulation. The dashed black line shows the fitted line on the $m=44$ mode.}
\label{random_start_1e4_Ek}
\end{figure}

The PIC simulations are performed with the codes EDIPIC, VSim, and XES1. EDIPIC is a locally developed code that uses the direct-implicit method to integrate the Vlasov--Poisson system of equations in a 1D3V (one spatial dimension and three velocity dimensions) geometry \cite{sydorenko_2006}. VSim is a commercial PIC  package that uses the VORPAl computation engine~\cite{nieter2004vorpal} to simulate plasmas. In addition, we perform some simulations with XES1~\cite{birdsall2004plasma}. For the simulations PIC1, PIC2, and PIC3, the calculation of the linear growth rates is done using $10^4$ macroparticles per cell. \Cref{random_start_1e4_ES} shows the evolution of the electrostatic energy in random-start PIC simulations. At $t=0$, the electrostatic energy of PIC simulations is very small because at this time the negative and positive charges are distributed uniformly in the system, so that the system is in a quasi-neutral state. This characteristic is embedded in all PIC simulation codes used in this study, independently of their initialization method. However, after $t=0$, the ES energy jumps to a finite value. This jump, which is absent in the Vlasov simulations, depends on the initial noise in the velocity space of PIC simulations. Therefore, the ES energy at the second collected time ($t=500\Delta t$) can be seen as a measure of the initial noise in the simulations. \Cref{random_start_1e4_ES} shows that, relative to the Vlasov simulations (\Cref{low_noise_ES}), the ES energy is much larger.  This indicates that the amount of initial noise in the PIC simulations is much larger than that of the Vlasov simulations. \Cref{random_start_1e4_Ek} shows the evolution of the chosen modes separately. The initial growth in these modes is essentially oscillatory instead of being linear, and therefore, the SE of the growth rate measurements is much larger than unity (\Cref{table:growth_numbers_2vte_PIC123}).  We can also define the 99\% confidence interval of the measured growth rates as $\gamma(1\, \pm\, 2.576\, \text{SE})$. The theoretical growth rates can be seen to not lie in the 99\% confidence interval of the fits, and thus the measured growth rates cannot be seen as equal to the theoretical growth rates to within the measurement error. We note that the applicability of the 99\% confidence interval, for this purpose, is limited to the simulations with significant noise in their linear growth regime (i.e.,~using the confidence interval is not meaningful in simulations where $SE\rightarrow 0$ because in such cases the confidence interval nearly vanishes). The average SNR of the chosen modes in the growth region is $-8.1$ dB in PIC1, $-13.75$ dB in PIC2, and $-15$ dB in PIC3.  Therefore, the EDIPIC code introduces the least noise power, and XES1 introduces the most noise power in the growth region of the three simulations. We will see that this trend of SNR also applies for the three codes in all other simulations of this study. The small SNR in all three simulations indicates the high noise power in the random-start PIC simulations.  To investigate the convergence with respect to the spatial resolution, we repeated the PIC2 simulation with 1024 and 4096 spatial grid points, and the level of noise and the reported results remained close to the original PIC2 simulation. In order to study the effect of only changing the spatial resolution, we note that it is important to not change the number density of macroparticles. The PIC2 simulation was also repeated with a doubled time step size, and again, no significant change was observed in the results.

\begin{table}[htbp]
\begin{tabular}[b]{|c|c|c|c|c|c|c|c|}
\hline
$k$  & \begin{tabular}[c]{@{}c@{}}$\gamma$ (Theory)\\ $\times 10^8 \text{ s}^{-1}$\end{tabular} & \begin{tabular}[c]{@{}c@{}}$\gamma$ (PIC1)\\ $\times 10^8 \text{ s}^{-1}$\end{tabular}& \begin{tabular}[c]{@{}c@{}}SE (PIC1)\\ \%\end{tabular} & \begin{tabular}[c]{@{}c@{}}$\gamma$ (PIC2)\\ $\times 10^8 \text{ s}^{-1}$\end{tabular}& \begin{tabular}[c]{@{}c@{}}SE (PIC2)\\ \%\end{tabular}& \begin{tabular}[c]{@{}c@{}}$\gamma$ (PIC3)\\ $\times 10^8 \text{ s}^{-1}$\end{tabular}& \begin{tabular}[c]{@{}c@{}}SE (PIC3)\\ \%\end{tabular}\\ \hline
30 & 0.90                                                                            & 2.79 & 5.80 & 2.96 & 1.83 & 3.37 & 1.72                                                                        \\ \hline
37 & 1.08                                                                            & 3.68 & 5.30 & 2.37 & 3.46 & 2.02 & 4.70                                                                         \\ \hline
44 & 1.17                                                                            & 3.00 & 6.00 & 3.54 & 1.46 & 3.44  & 1.98                                                                       \\ \hline
51 & 1.07                                                                            & 4.03 & 3.00 & 2.76 & 4.14 & 2.38 & 2.61                                                                    \\ \hline
\hline
Average & 1.06 & 3.38 & 5.02 & 2.91 & 2.72 & 2.80 & 2.75 \\
\hline
\end{tabular}
\caption{Comparison of the theoretical growth rates with the growth rates observed in the PIC1, PIC2, and PIC3 simulations\label{table:growth_numbers_2vte_PIC123}.}
\end{table}

\begin{figure}[htbp]
\centering
\captionsetup[subfigure]{labelformat=empty}
\subcaptionbox{\label{EdiPic_100kmp_2vte_EsPe-PIC4}}{\includegraphics[width=.49\linewidth]{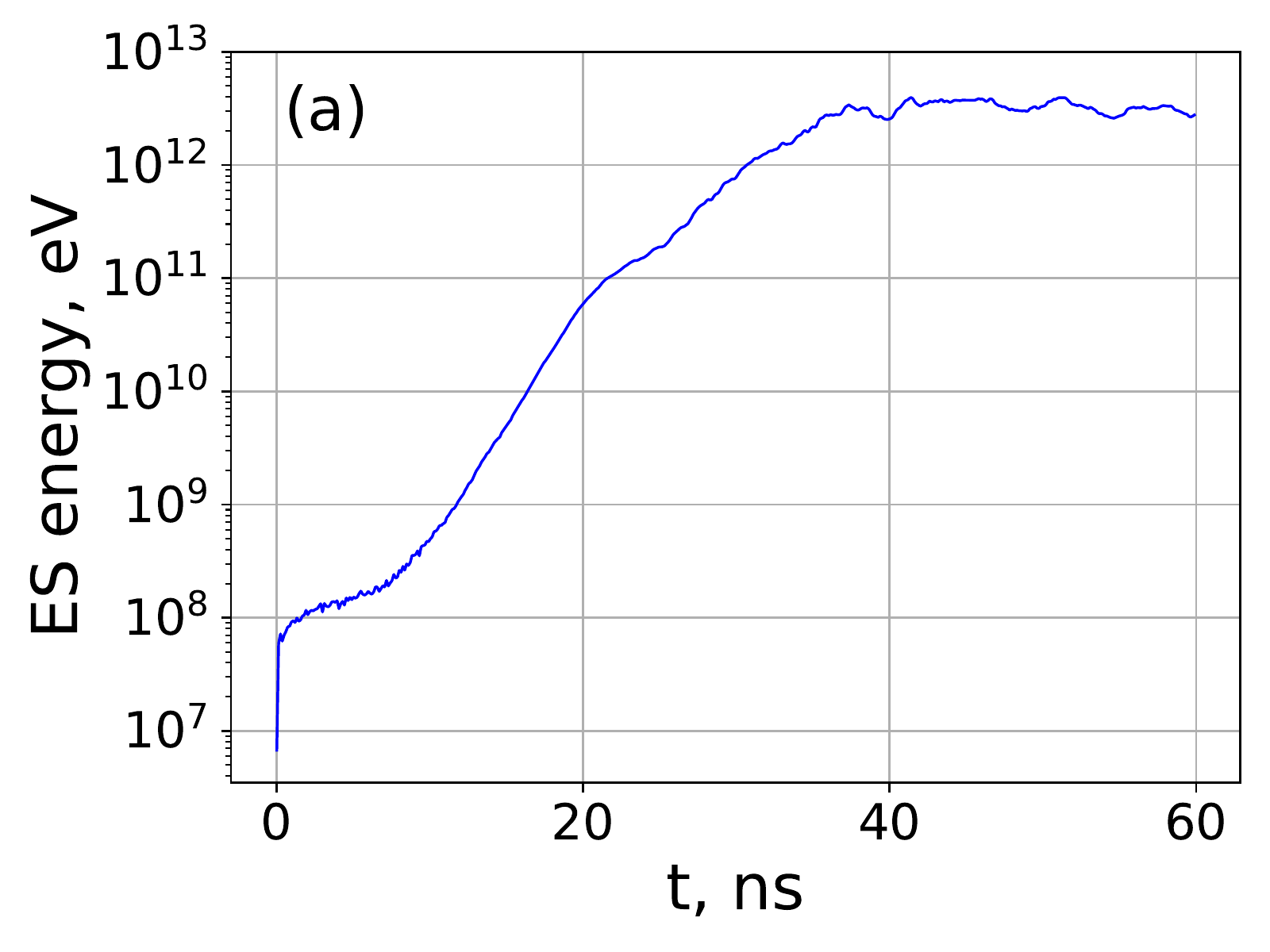}}
\subcaptionbox{\label{es_100000}}{\includegraphics[width=.49\linewidth]{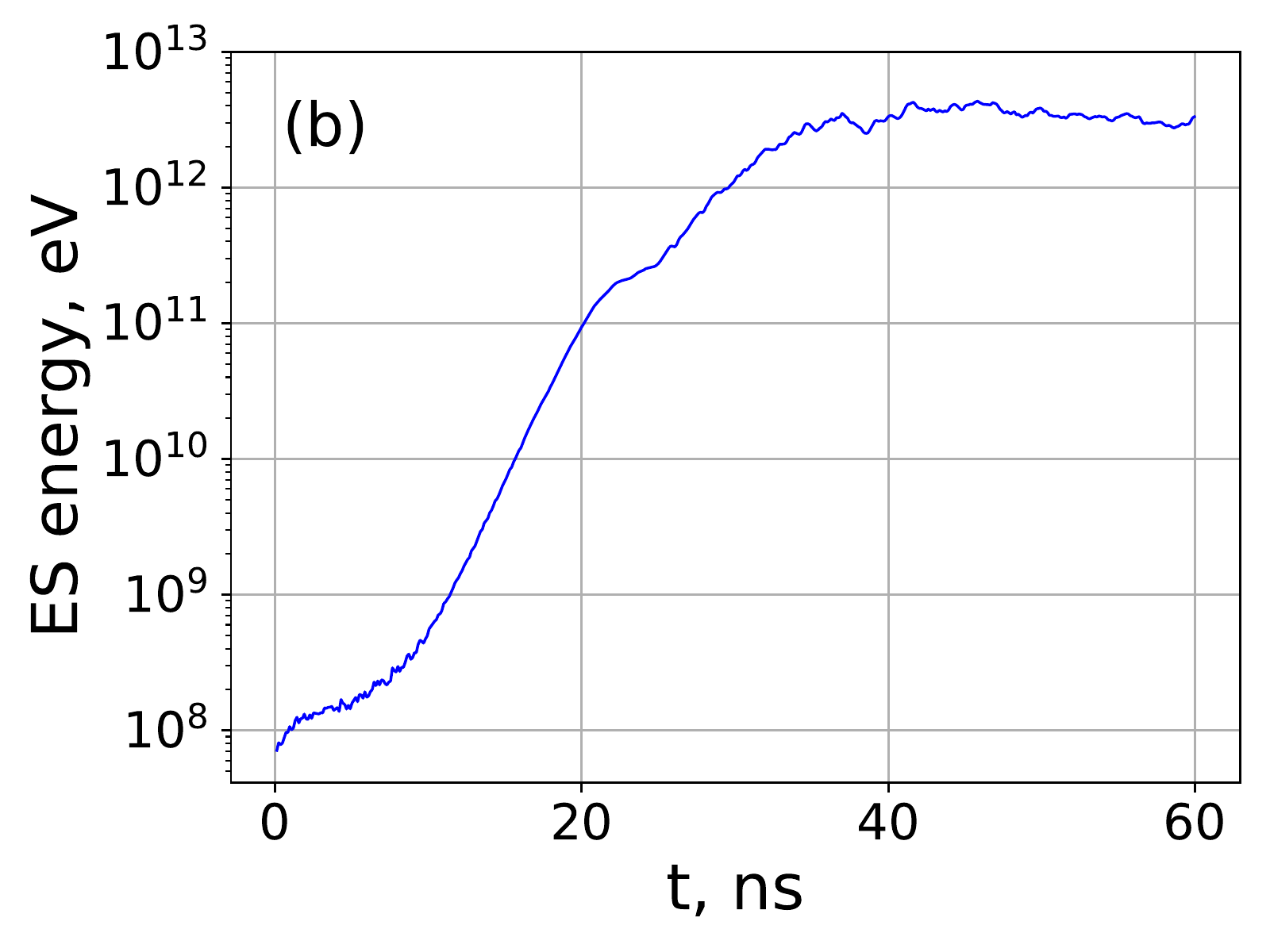}}
\caption{The evolution of the electrostatic energy in random-start PIC using $10^5$ macroparticles per cell from a) EDIPIC (PIC4) b) VSim (PIC5) simulations.}
\label{random_start_1e5_ES}
\end{figure}

\begin{figure}[htbp]
\centering
\captionsetup[subfigure]{labelformat=empty}
\subcaptionbox{\label{EdiPic_100kmp_2vte_Ex_Growth-PIC4}}{\includegraphics[width=.49\linewidth]{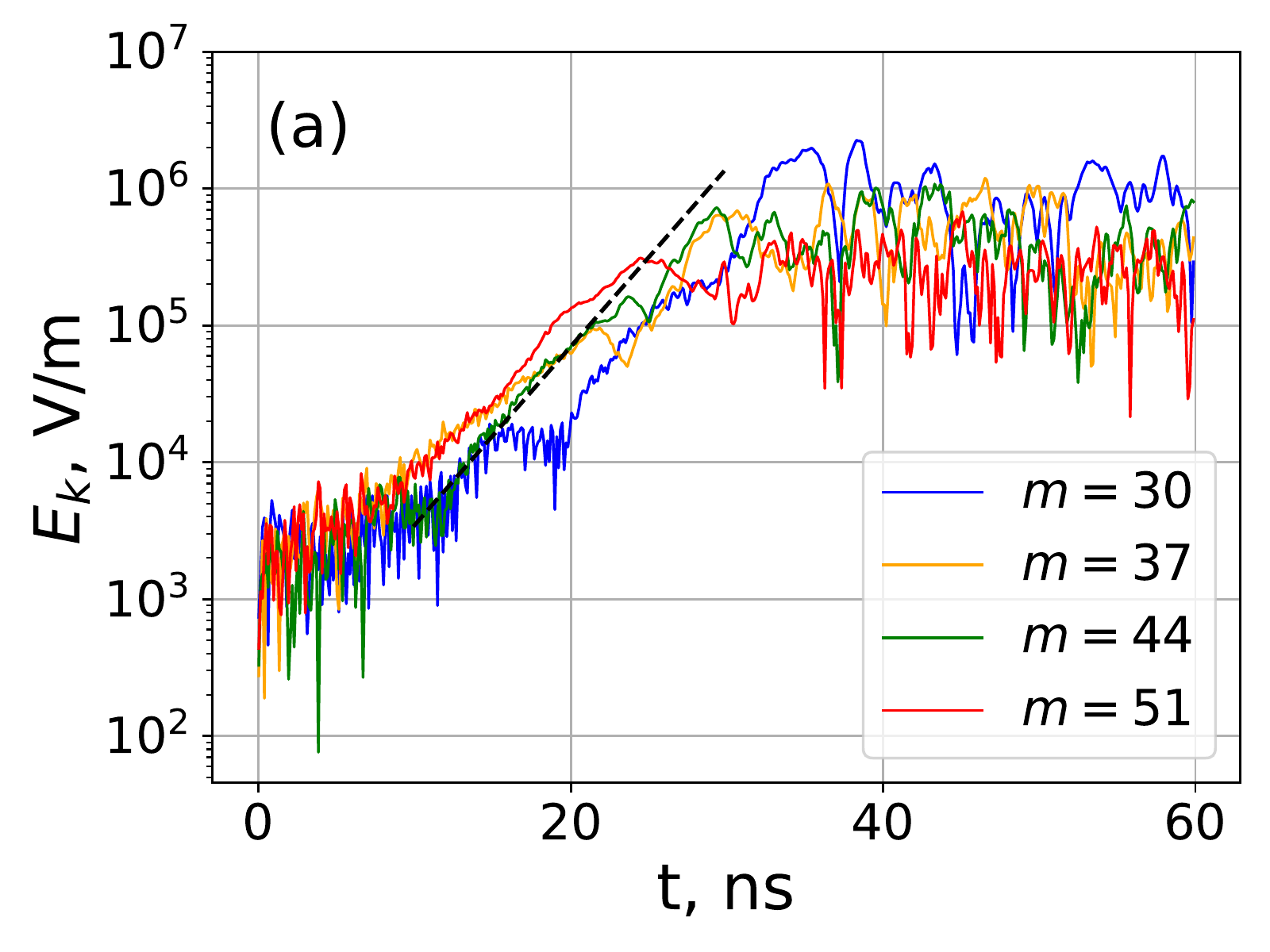}}
\subcaptionbox{\label{E_k_growth_100000}}{\includegraphics[width=.49\linewidth]{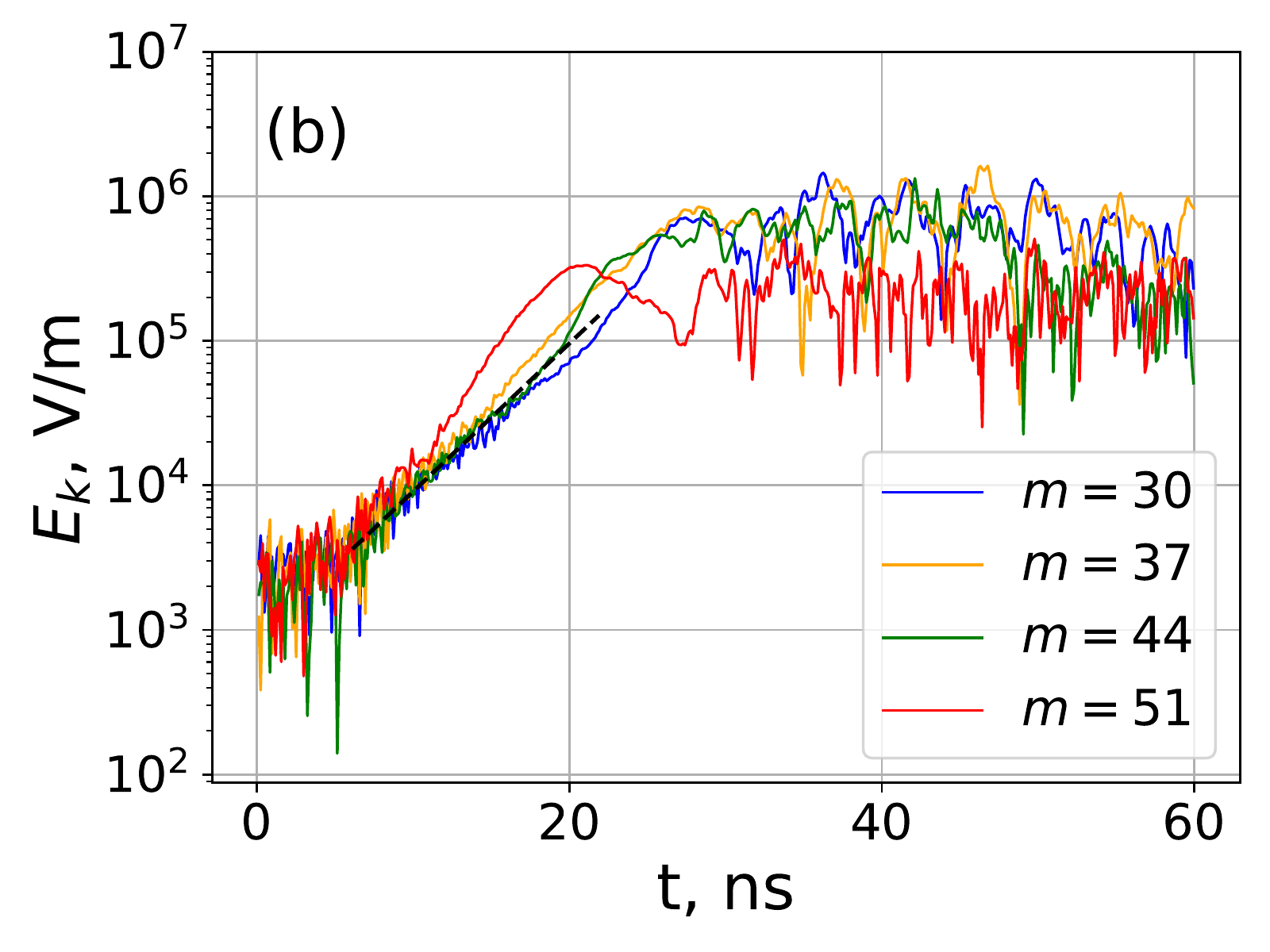}}

\caption{The evolution of individual modes of the electric field in random-start PIC simulations using $10^5$ macroparticles per cell from a) EDIPIC (PIC4), b) VSim (PIC5) simulations. The dashed black line shows the fitted line on the $m=44$ mode.}
\label{random_start_1e5_Ek}
\end{figure}

The inaccurate growth rates of the PIC simulations suggest that the noise level in these simulations is so high that it severely influences with the linear growth. Therefore, to reduce the statistical noise level, we increase the number of macroparticles per cell to $10^5$ and redo the PIC simulations (PIC4 and PIC5). The initial electrostatic energy in this case is reduced by an approximate factor of $1/10$, whereas the initial amplitude of individual modes is reduced by an approximate factor of $1/\sqrt{10}$ (compare \Cref{random_start_1e5_ES} and \Cref{random_start_1e5_Ek} with \Cref{random_start_1e4_ES} and \Cref{random_start_1e4_Ek}, respectively). This indicates that the initial noise is reduced approximately by a factor of $1/\sqrt{N_p}$, as expected. The measured growth rates for the random-start PIC simulations with $10^5$ particle per cell is reported in \Cref{table:growth_numbers_2vte_PIC45}. The growth rates of PIC4 simulation with $10^5$ macroparticles per cell are smaller than their counterparts in PIC1 with $10^4$ macroparticles per cell. Accordingly, they are closer to the theoretical growth rates. On the other hand, we see a reduction in spurious oscillation in the linear regime, so that the SEs of the PIC4 simulation are less than those of the PIC1 simulation. In \Cref{table:growth_numbers_2vte_PIC45}, we can also see that the average growth rate in the PIC5 simulation is closer to the theory than its corresponding PIC2 simulation. However, in a few modes, such as $m=51$, we see that the measured growth rate in PIC5 is farther from the theory than it is in PIC2. In both PIC4 and PIC5, the measured linear growth rates are still much larger than the theoretical growth rates. The average SNR of the chosen modes is $-15.58$ dB in PIC4 and $-14.75$ dB in PIC5.

\begin{table}[htbp]
\begin{tabular}[b]{|c|c|c|c|c|c|}
\hline
$m$  & \begin{tabular}[c]{@{}c@{}}$\gamma$ (Theory)\\ $\times 10^8 \text{s}^{-1}$\end{tabular} & \begin{tabular}[c]{@{}c@{}}$\gamma$ (PIC4)\\ $\times 10^8 \text{s}^{-1}$\end{tabular}& \begin{tabular}[c]{@{}c@{}}SE (PIC4)\\ \%\end{tabular}  &\begin{tabular}[c]{@{}c@{}}$\gamma$ (PIC5)\\ $\times 10^8 \text{s}^{-1}$\end{tabular} &\begin{tabular}[c]{@{}c@{}}SE (PIC5)\\ \%\end{tabular} \\ \hline
30 & 0.90                                                                            & 2.45 & 1.60& 2.08 & 2.23                                                                   \\ \hline
37 & 1.08                                                                            & 1.81 & 1.71 & 2.56 & 2.01                                                                       \\ \hline
44 & 1.17                                                                            & 3.00 & 1.90 & 2.55 & 2.74                                                                        \\ \hline
51 & 1.07                                                                            & 2.69 & 1.11 & 3.29 & 1.17                                                                        \\ \hline
\hline
Average & 1.06 & 2.49 & 1.58 & 2.62 & 2.04  \\
\hline
\end{tabular}
\caption{The comparison of the theoretical growth rates with the growth rates observed in the PIC4 and PIC5 simulations.\label{table:growth_numbers_2vte_PIC45}}
\end{table}

\begin{figure}[htbp]
\centering
\captionsetup[subfigure]{labelformat=empty}
\subcaptionbox{\label{t_Ek_correct-VL3}}{\includegraphics[width=.46\linewidth]{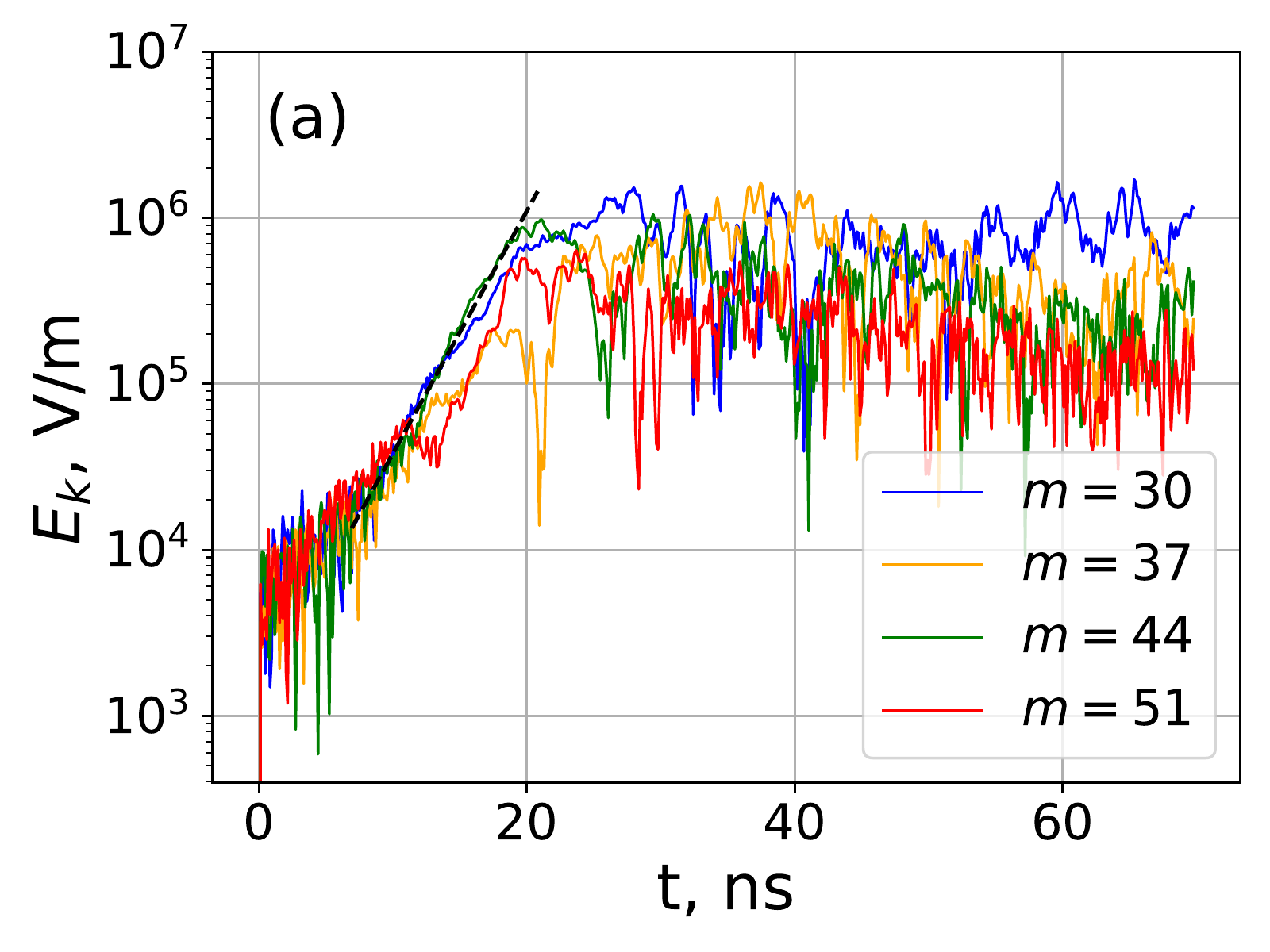}}
\subcaptionbox{\label{t_ES-VL3}}{\includegraphics[width=.46\linewidth]{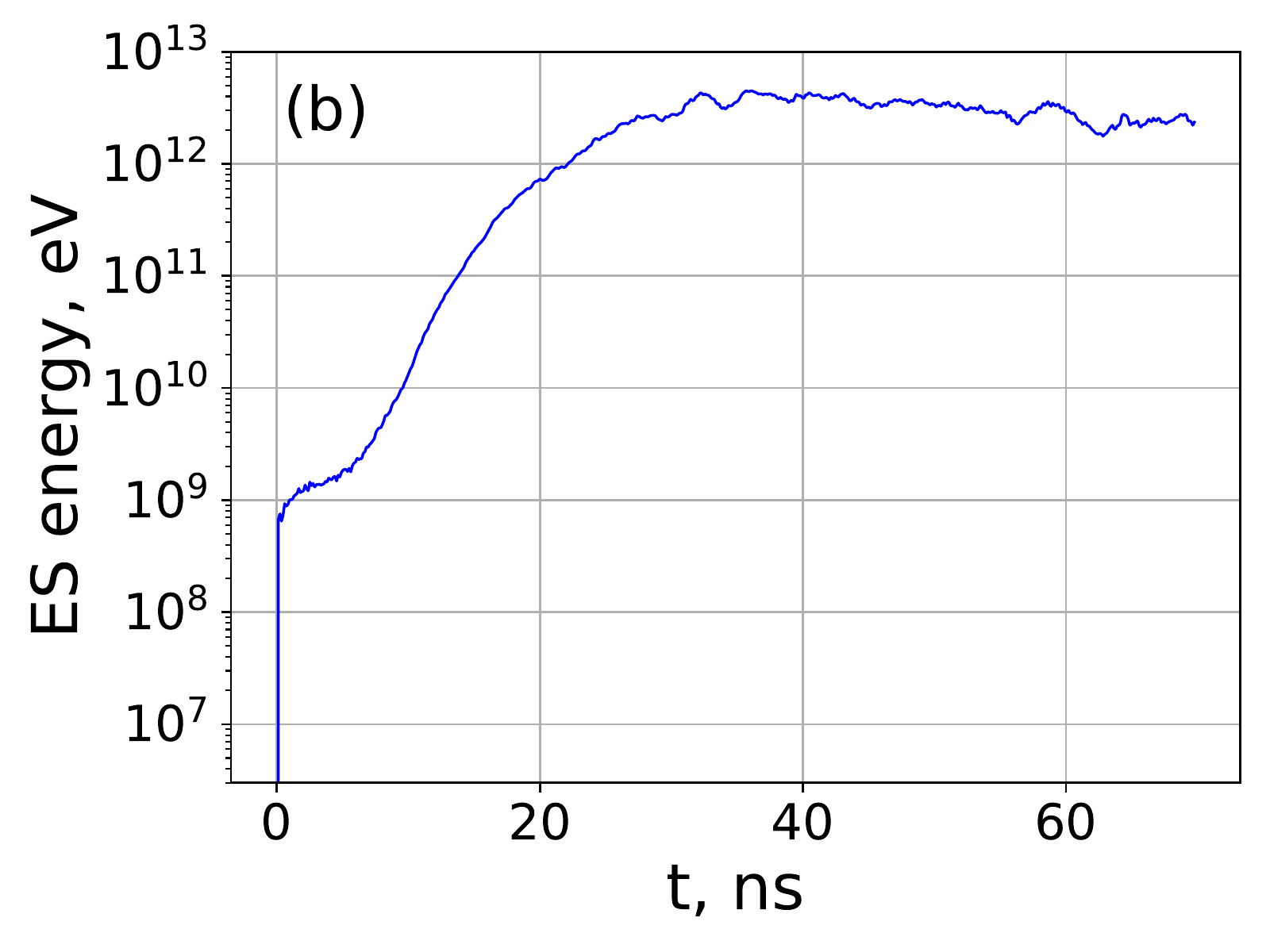}}
\caption{a) The evolution of individual modes of the  electric field. The dashed black line shows the fitted line on the $m=44$ mode. b) The evolution of the electrostatic energy. Both figures are from the semi-Lagrangian code (VL3) with the initial condition  taken from PIC2 simulation.}
\end{figure}

To investigate the problem of inaccurate growth rates in random-start PIC simulations, we introduce a test simulation with the semi-Lagrangian Vlasov code. In this simulation (VL3), we tabulate the initial condition of macroparticles in PIC2 simulation to find the corresponding distribution function and use it as the initial condition for the semi-Lagrangian Vlasov code. By doing this, we introduce the same initial noise as the PIC simulations into the Vlasov simulation. We then repeat the ES energy and  mode growth rate analyses using the results of the VL3 simulation (\Cref{t_ES-VL3,t_Ek_correct-VL3}). As with the PIC simulations, we see that the growth of the chosen modes is oscillatory, and the resultant growth rates are much larger than those predicted from theory (\Cref{table:growth_numbers_2vte_VL3}). This strongly suggests that the influence of the initial noise in the PIC simulations is the cause of inaccurate growth rates. 

\begin{table}[htbp]
\begin{tabular}[b]{|c|c|c|c|}
\hline
$m$  & \begin{tabular}[c]{@{}c@{}}$\gamma$ (Theory)\\ $\times 10^8$ s$^{-1}$\end{tabular} & \begin{tabular}[c]{@{}c@{}}$\gamma$ (VL3)\\ $\times 10^8$ s$^{-1}$\end{tabular}&\begin{tabular}[c]{@{}c@{}}SE (VL3)\\ \%\end{tabular} \\ \hline
30 & 0.90                                                                            & 3.00 &  1.57                \\ \hline
37 & 1.08                                                                            & 2.30 &  2.71                                                                   \\ \hline
44 & 1.17                                                                            & 3.50  &  1.23                                                                     \\ \hline
51 & 1.07                                                                            & 2.17 &  2.99                                                                     \\ \hline
\hline
Average & 1.06 & 2.74 & 2.13 \\ 
\hline
\end{tabular}
\caption{Comparison of the theoretical growth rates with the growth rates observed in the VL3 simulation.\label{table:growth_numbers_2vte_VL3}}
\end{table}

\section{The effect of a small flattening of electron distribution function on linear growth rates}\label{sec:small_flattening}

\begin{figure}[htbp]
\centering
\includegraphics[width=0.6\linewidth]{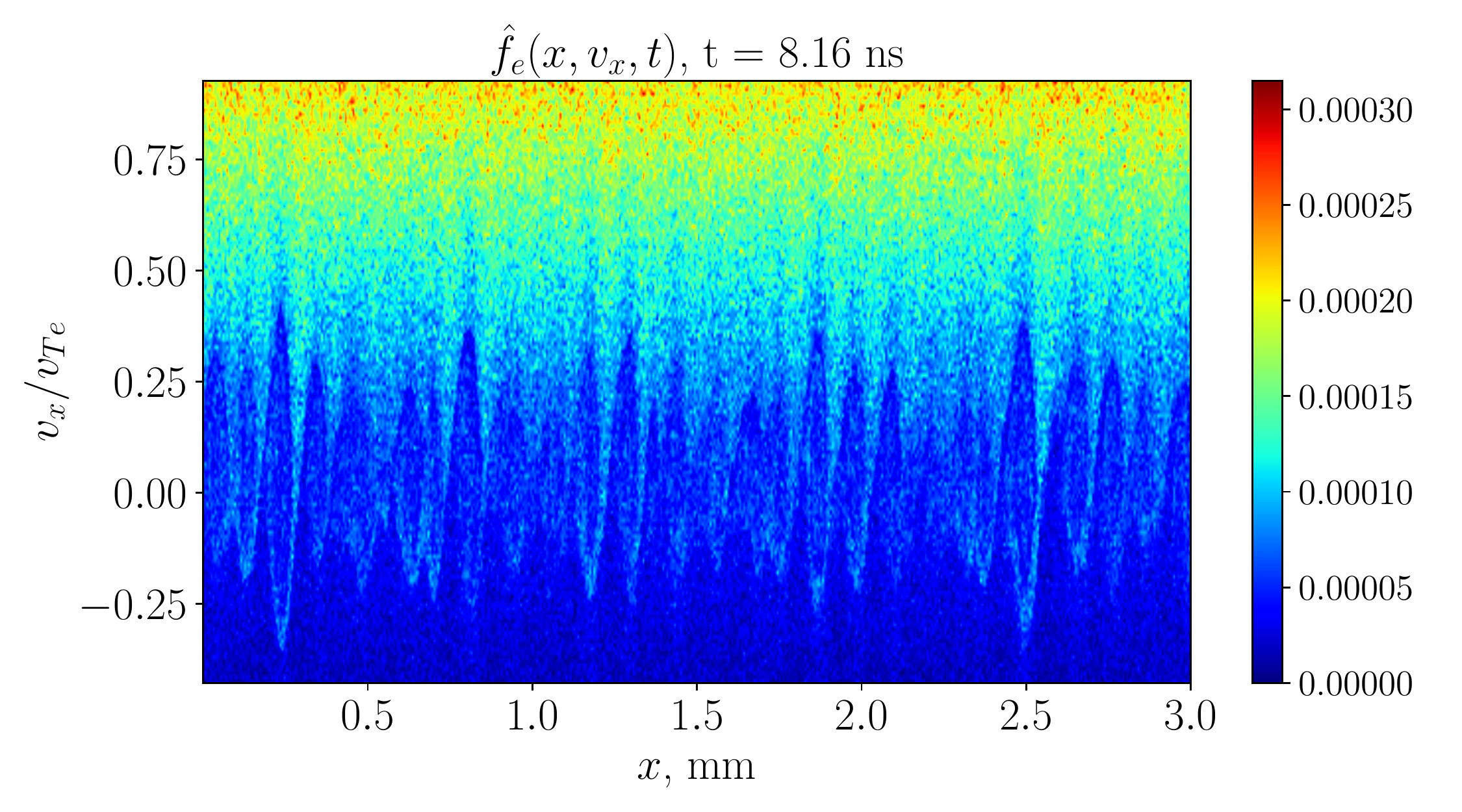}
\caption{The holes in the electron distribution function at $t=8.16$ ns of simulation PIC3.}
\label{fe_8ns_halfx_xes1_rand_1e4}
\end{figure}

\begin{figure}[htbp]
\centering
\captionsetup[subfigure]{labelformat=empty}
\subcaptionbox{\label{evdf_pic_6mm_lin_npc10000}}{\includegraphics[width=.46\linewidth]{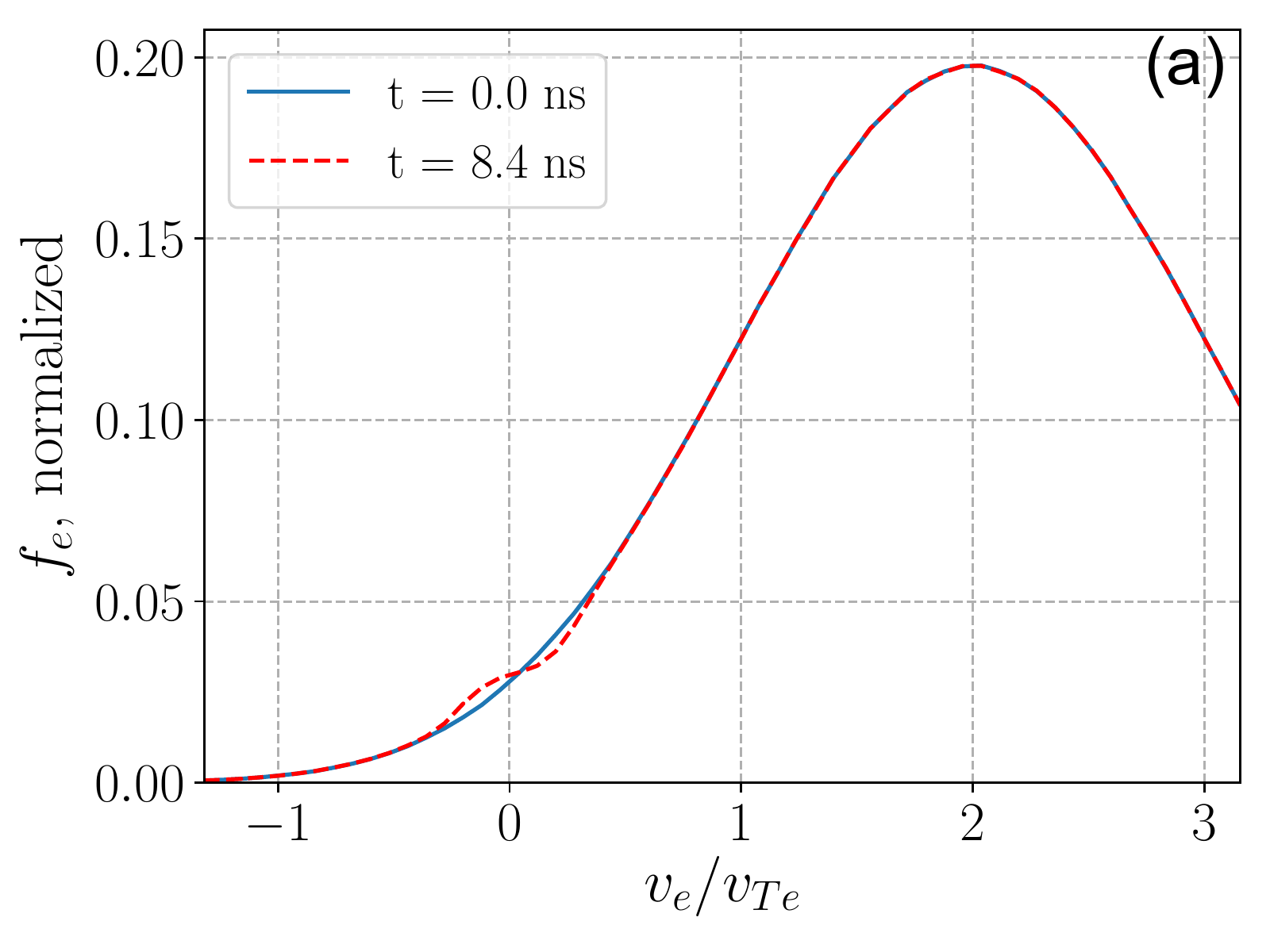}}
\subcaptionbox{\label{f_maxw_modified}}{\includegraphics[width=.46\linewidth]{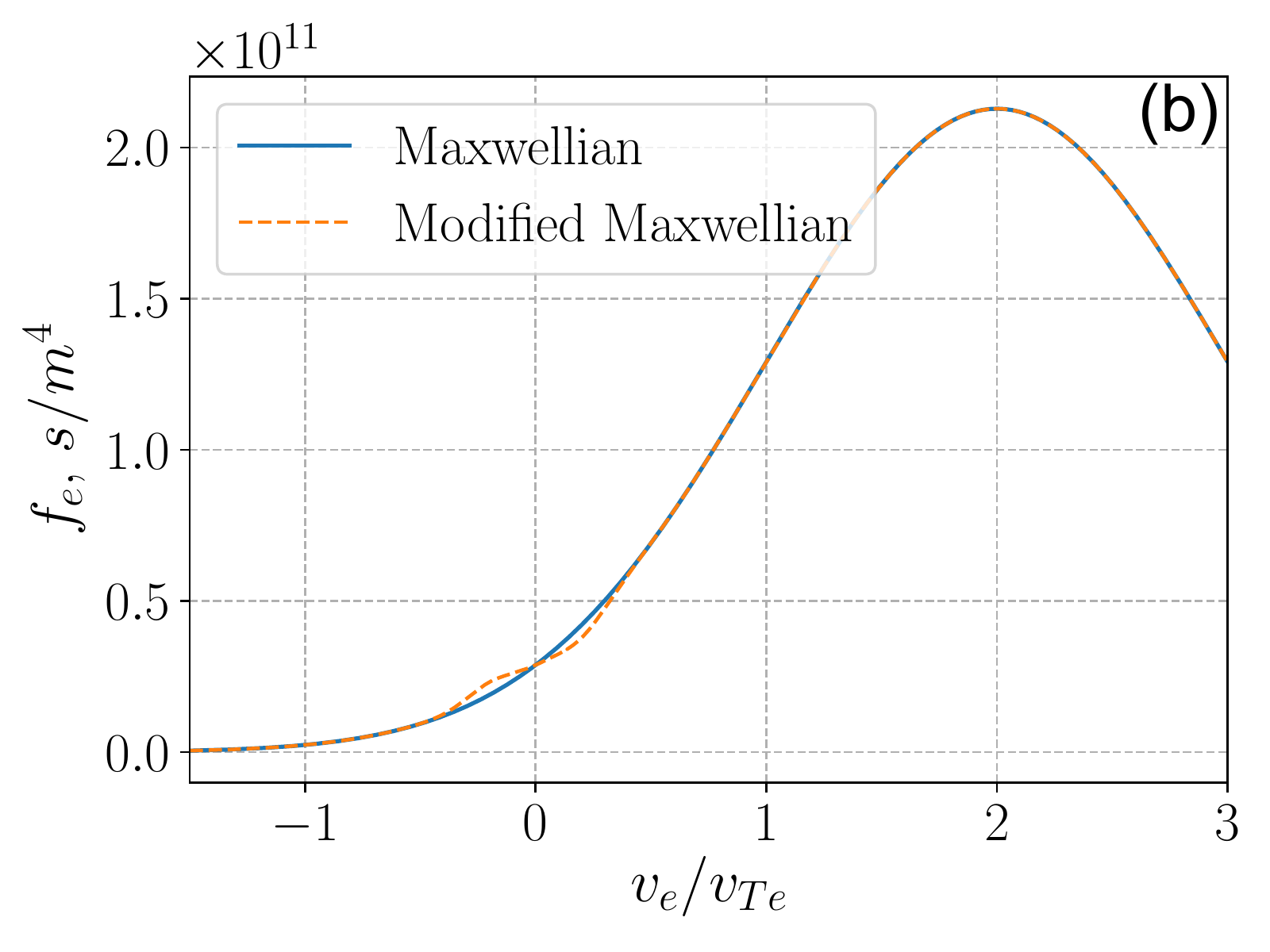}}
\caption{a) Electron VDF in PIC3 simulation at $x=L/2$. b) Electron VDF in \Cref{flattened_VDF} for $\alpha=0.002$, $v_{te}' = 0.1 v_{te}$, and $v_0' = 0.1 v_0$. For comparison, the Maxwellian VDF is also shown in blue, in each figure.\label{mod_fve_and_Maxwellian}}
\end{figure}

In \Cref{fe_8ns_halfx_xes1_rand_1e4}, the coherent structures (holes) in the electron velocity distribution function (VDF) are shown. These structures appear early in the  PIC3 simulation (similar structures are observed in other PIC simulations with random-start and VL3) and are a result of trapping of electrons and reflect a small flattening in their Maxwellian velocity distribution function (\Cref{evdf_pic_6mm_lin_npc10000}). This flattening is in fact a depletion of the electrons in the positive velocity region of electron VDF that leads to an increase in electrons in the negative velocity region. To model the flattened velocity distribution function, we add and subtract two shifted Maxwellians (beams) from the initial electron VDF of \Cref{electron_init_VDF}:
\begin{gather}\label{flattened_VDF}
f_m(v)=\frac{n_0}{\sqrt{2\pi}v_{te}}\exp(-\frac{(v-v_0)^2}{2v_{te}^2})+\frac{\alpha n_0}{\sqrt{2\pi}v^{\prime}_{te}}\exp(-\frac{(v+v^{\prime}_0)^2}{2v^{\prime 2}_{te}})-
\frac{\alpha n_0}{\sqrt{2\pi}v^{\prime}_{te}}\exp(-\frac{(v-v^{\prime}_0)^2}{2v^{\prime 2}_{te}}),
\end{gather}
where $v^{\prime}_0$ is the drift velocity of the added beams, $v^{\prime}_{te}$ is their thermal velocity, and $\alpha$ is their density fraction. To replicate the flattened electron VDF in the simulations, we take $\alpha=0.002$, $v_{te}' = 0.1\, v_{te}$, and $v_0' = 0.1\, v_0$. \Cref{f_maxw_modified} shows this VDF and compares it with the Maxwellian VDF ($\alpha=0$). Using this VDF, the linear desperation relation reads:
\begin{gather} \label{disp_kinetic_mod}
     1 +\frac{1}{k^2 \lambda_{Di}^2} \mathrm{Z^{\prime}} \left( \frac{\omega}{\sqrt{2} k v_{ti}} \right)+ \frac{1}{k^2 \lambda_{De}^2} \mathrm{Z^{\prime}} \left( \frac{\omega - kv_0}{\sqrt{2} k v_{te}} \right) + 
    \alpha \frac{1}{k^2 \lambda_{De}^2} \mathrm{Z^{\prime}} \left( \frac{\omega+kv_0'}{\sqrt{2} k v_{te}'} \right) - \\ \nonumber 
     \alpha \frac{1}{k^2 \lambda_{De}^2} \mathrm{Z^{\prime}} \left( \frac{\omega-kv^{\prime}_0}{\sqrt{2} k v^{\prime}_{te}} \right) = 0,
\end{gather}
where $\lambda_{Di,De}$ are the ion and electron Debye lengths.

\begin{figure}[htbp]
\centering
\includegraphics[width=0.6\linewidth]{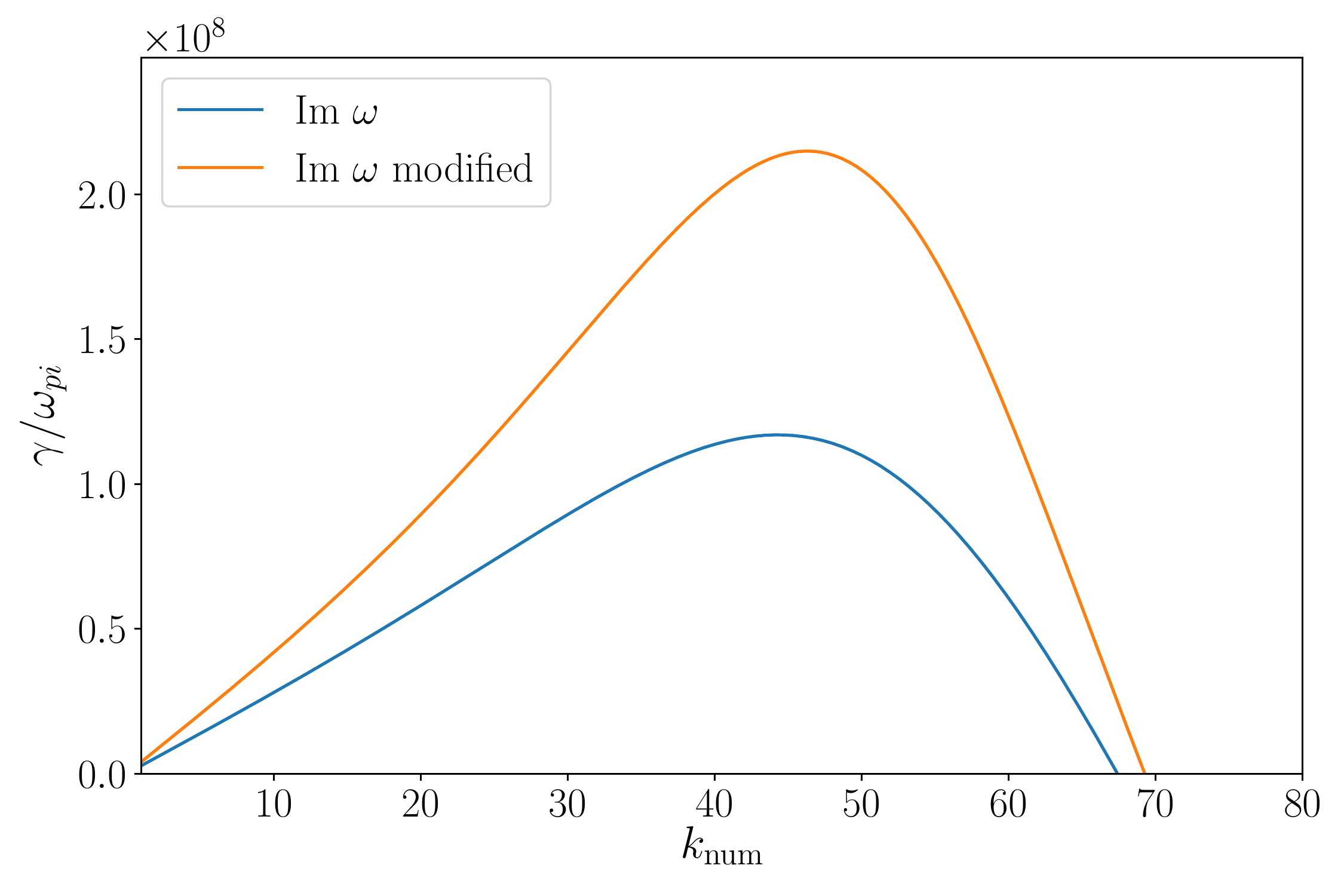}
\caption{Growth rate from the modified dispersion equation~(\Cref{disp_kinetic_mod}), with $\alpha=0.002$, $v_{te}' = 0.1 v_{te}$, and $v_0' = 0.1 v_0$. For comparison, the growth rates of original dispersion relation are also shown.}
\label{kinetic_disp_2str6mm_im_re_mod}
\end{figure}

Solving this dispersion relation, we find the growth rates as shown in \Cref{kinetic_disp_2str6mm_im_re_mod}. We see that the small flattening in the electron Maxwellian VDF leads to much larger growth rates.

\section{Using quiet-start initialization to reduce the effect of noise in PIC simulations}\label{sec:quite_start}

In this section, we report on several PIC simulations (PIC6 to PIC12) that use the quiet-start initialization. The quiet-start initialization, proposed by J. A. Byers \cite{byers1970perpendicularly}, employs a smooth loading of macroparticles in phase space to reduce the noise in PIC simulation. In this method, the initial placement of macroparticles in $x$-$v$ space starts with desired space and velocity densities, $n_0(x)$  and $f_0(v)$, respectively. The method for generating the positions and velocities of each particle from density functions is based on inversion of the ``cumulative density",
\begin{equation}
R_{s}(\xi)=\frac{\int_{a}^{\xi}d(\xi^{\prime})d\xi^{\prime}}{\int_{a}^{b } d(\xi^{\prime})d\xi^{\prime}},
\end{equation}
where $d(\xi')$ is the density function and $\xi$ can be either $x$ or $v$. This cumulative density calculates the cumulative probability in each component $x$ or $v$. $R_s$ can be a uniform set of numbers or a numerical sequence that generates quasi-random numbers with low discrepancy. Several sequences have been proposed in the literature, including the bit-reversed (or Hammersley) sequence\cite{hammersley1964monte,birdsall2004plasma,gonichon1993quiet}, Sobol sequence\cite{bratley1988algorithm}, and Fibonacci sequence\cite{parker1993fully}. The inversion of the $R_s$ function, by either analytical or numerical means, produces the position or velocity of macroparticles. The $R_{s}$ set for velocity and position should be uncorrelated to avoid unwanted bunching in phase space. The quiet-start used in our PIC simulations utilizes the bit-reversed set for assigning particle positions and a uniform set of numbers for $R_s$ for particle velocity. This method of quiet-start has been described and implemented in Ref.~\onlinecite{birdsall2004plasma}. In  practice, a particular mode is perturbed with a finite amplitude initially.

\begin{figure}[htbp]
\centering
\captionsetup[subfigure]{labelformat=empty}
\subcaptionbox{\label{EdiPic_10kmp_2vte_Ex_Growth-PIC6}}{\includegraphics[width=.49\linewidth]{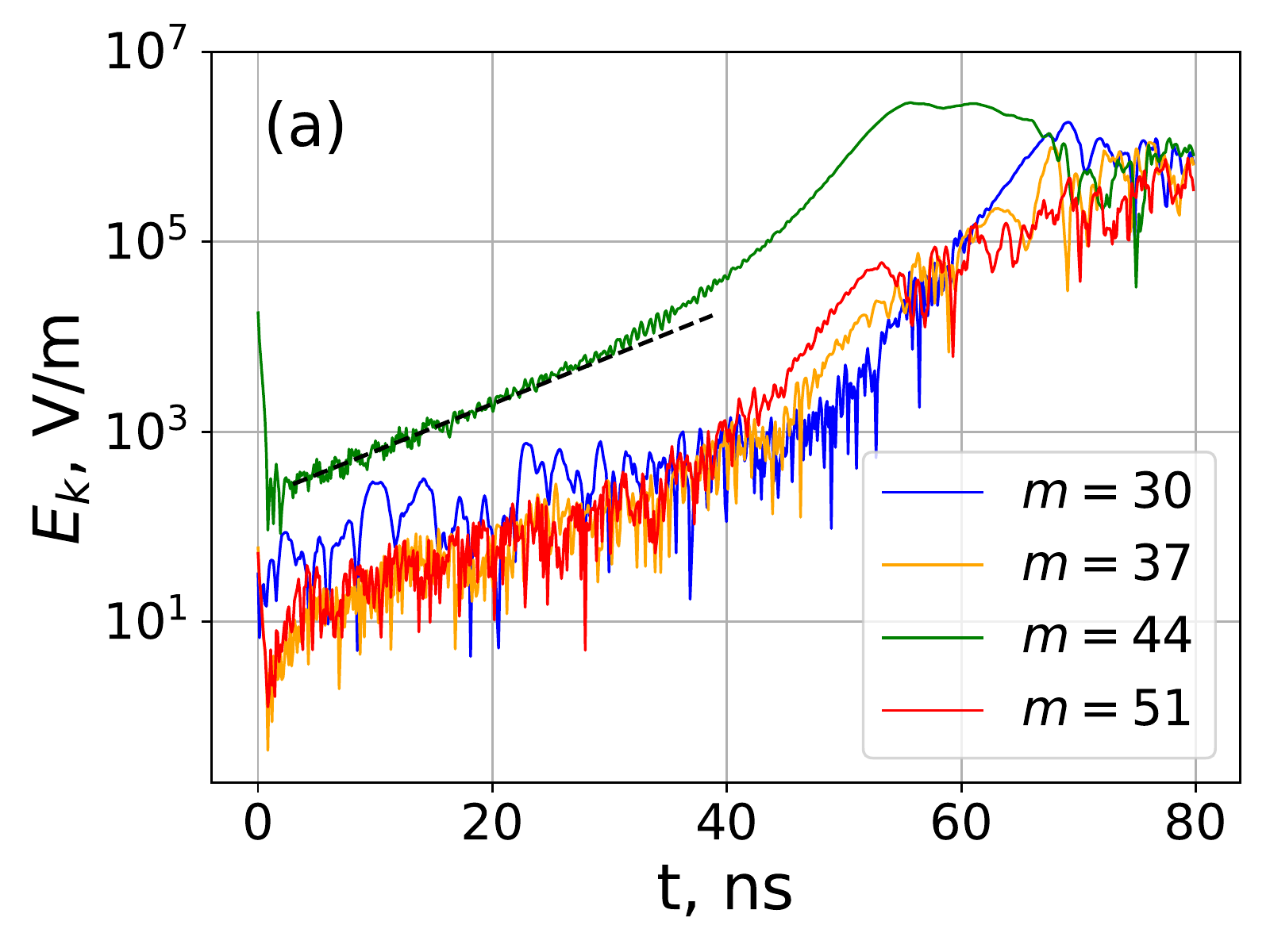}}
\subcaptionbox{\label{V-R-qs-20480k-M441e-8}}{\includegraphics[width=.49\linewidth]{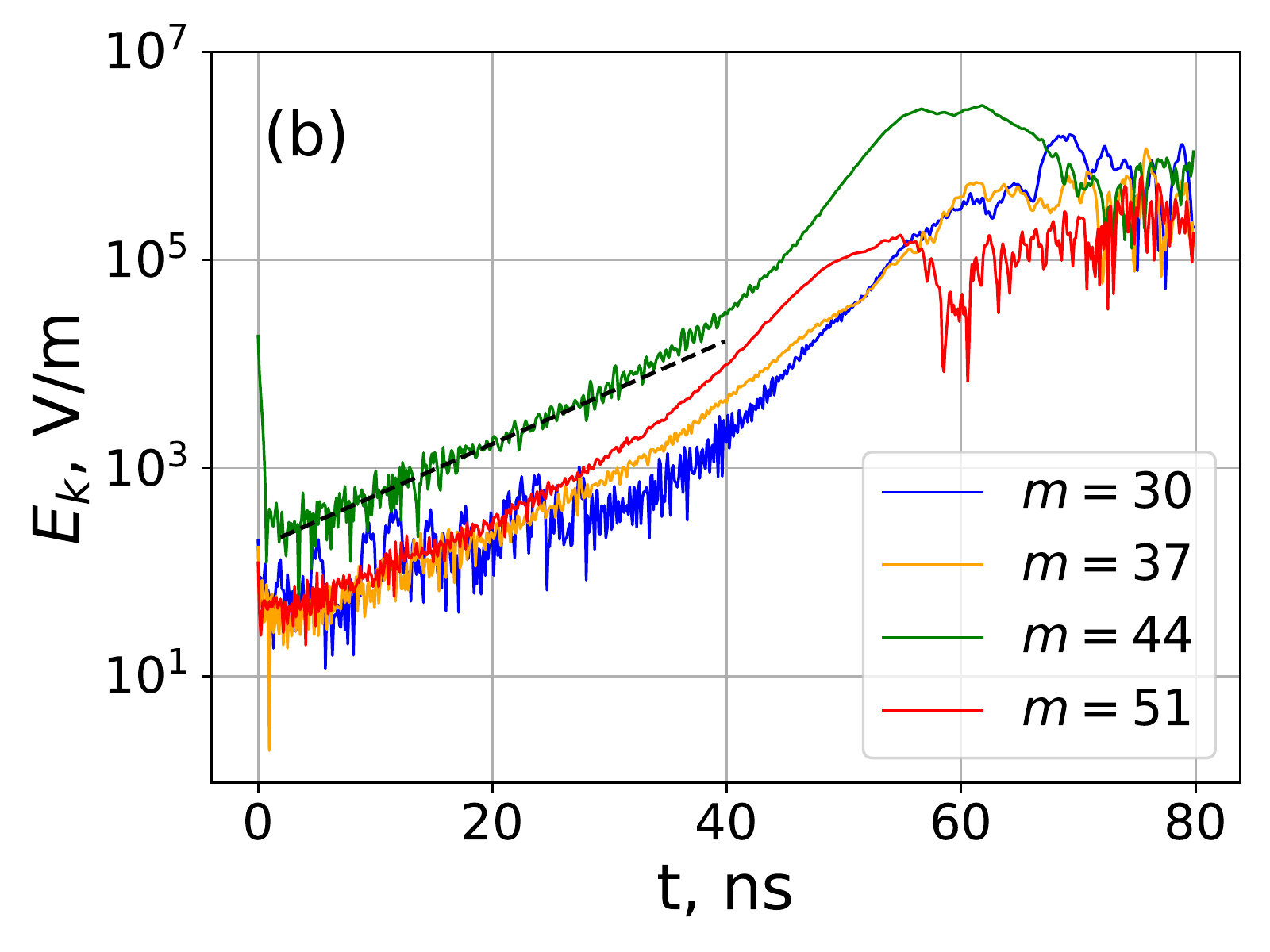}}

\subcaptionbox{\label{X-qs-M44-10kppc}}{\includegraphics[width=.49\linewidth]{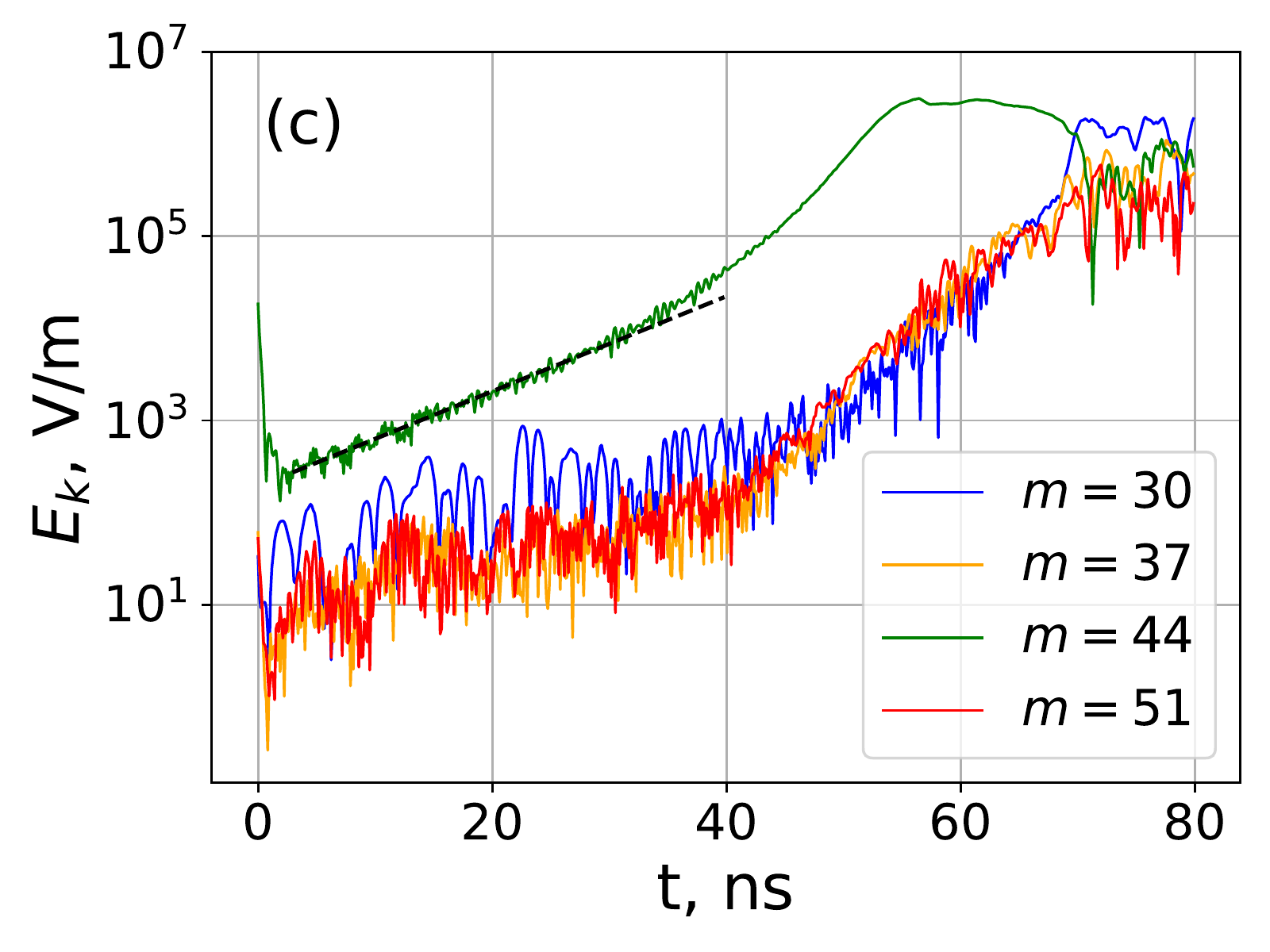}}
\caption{The evolution of individual modes of the  electric field in quiet-start PIC simulations, using $10^4$ macroparticles per cell, from a) EDIPIC (PIC6)  b) VSim (PIC7) c) XES1 (PIC8) simulation. The dashed black line shows the fitted line on the $m=44$ mode.}
\label{quiet_start_44_Ek}
\end{figure}

\Cref{quiet_start_44_Ek} shows the growth of individual modes in simulations PIC6, PIC7, and PIC8 using the quiet-start initialization. In these simulations, we have only perturbed the maximum growth rate mode $m=44$ initially. The growth rates measured by these simulations are reported in \Cref{table:growth_numbers_2vte_PIC678}. An obvious improvement, in comparison with the corresponding random-start PIC simulations (PIC1, PIC2, and PIC3), is that here the growth rate of the perturbed mode $m=44$ is close to its theoretical value in all three simulations. The average growth rates measured in  PIC6 (EDIPIC) and PIC7 (VSim) are also the same as the theoretical growth rates within the 99\% confidence interval. However, for the PIC8 simulation (XES1), the theoretical average growth rate is not in the 99\% confidence interval of the measured growth rate, and therefore, the two growth rates cannot be considered to be equal by this measure. This discrepancy is due to the inaccuracy of the growth rates for $m=\{30,37,51\}$ produced in the PIC8 simulation. In particular, the individual mode $m=30$ in all three PIC simulations is far from the theoretical value. The average SE in the PIC6 simulation is improved in comparison with its corresponding random-start simulation (PIC1). In contrast, the average SE of the modes is larger in PIC7 and PIC8 than in the corresponding random-start PIC simulations (PIC2 and PIC3, respectively). This indicates that, in general, the linearity of the growth has deteriorated in PIC7 and PIC8 simulations. The average SNR in the growth rate of chosen modes is $-26.5$ dB in PIC6, $-27.5$ dB in PIC7, and $-29.75$ dB in PIC8. These values of SNR are much smaller than what is reported in \cref{sec:low_v0} for the corresponding random-start PIC simulations. This is likely because of the high-frequency oscillations observed in the growth region of the quiet-start simulations (\Cref{quiet_start_44_Ek}) carry a large power of noise. 

\begin{table}[htbp]
\begin{tabular}[b]{|c|c|c|c|c|c|c|c|}
\hline
$m$  & \begin{tabular}[c]{@{}c@{}}$\gamma$ (Theory)\\ $\times 10^8s^{-1}$\end{tabular} & \begin{tabular}[c]{@{}c@{}}$\gamma$ (PIC6)\\ $\times 10^8s^{-1}$\end{tabular}& \begin{tabular}[c]{@{}c@{}}SE (PIC6)\\ \%\end{tabular} &\begin{tabular}[c]{@{}c@{}}$\gamma$ (PIC7)\\ $\times 10^8s^{-1}$\end{tabular}& \begin{tabular}[c]{@{}c@{}}SE (PIC7)\\ \%\end{tabular}&\begin{tabular}[c]{@{}c@{}}$\gamma$ (PIC8)\\ $\times 10^8s^{-1}$\end{tabular}& \begin{tabular}[c]{@{}c@{}}SE (PIC8)\\ \%\end{tabular} \\ \hline
30 & 0.90                                                                            & 0.61 & 5.61 & 0.77 & 8.36 & 0.66 &  7.28                                                               \\ \hline
37 & 1.08                                                                            & 1.05 & 2.52 & 1.14 & 2.18 & 0.61 &   5.91                                                                     \\ \hline
44 & 1.17                                                                            & 1.15 & 2.70 & 1.17 & 2.21 & 1.18 &   1.15                                                                     \\ \hline
51 & 1.07                                                                            & 1.12 & 2.62 & 1.23 & 1.43 & 0.62 &  6.21                                                                     \\ \hline
\hline
Average & 1.06 & 0.98 & 3.36 & 1.08 & 3.55 & 0.77 & 5.14 \\
\hline
\end{tabular}
\caption{The comparison of the theoretical growth rates with the growth rates observed in PIC6, PIC7, and PIC8 simulations.\label{table:growth_numbers_2vte_PIC678}}
\end{table}

\begin{figure}[htbp]
\centering
\includegraphics[width=.46\linewidth]{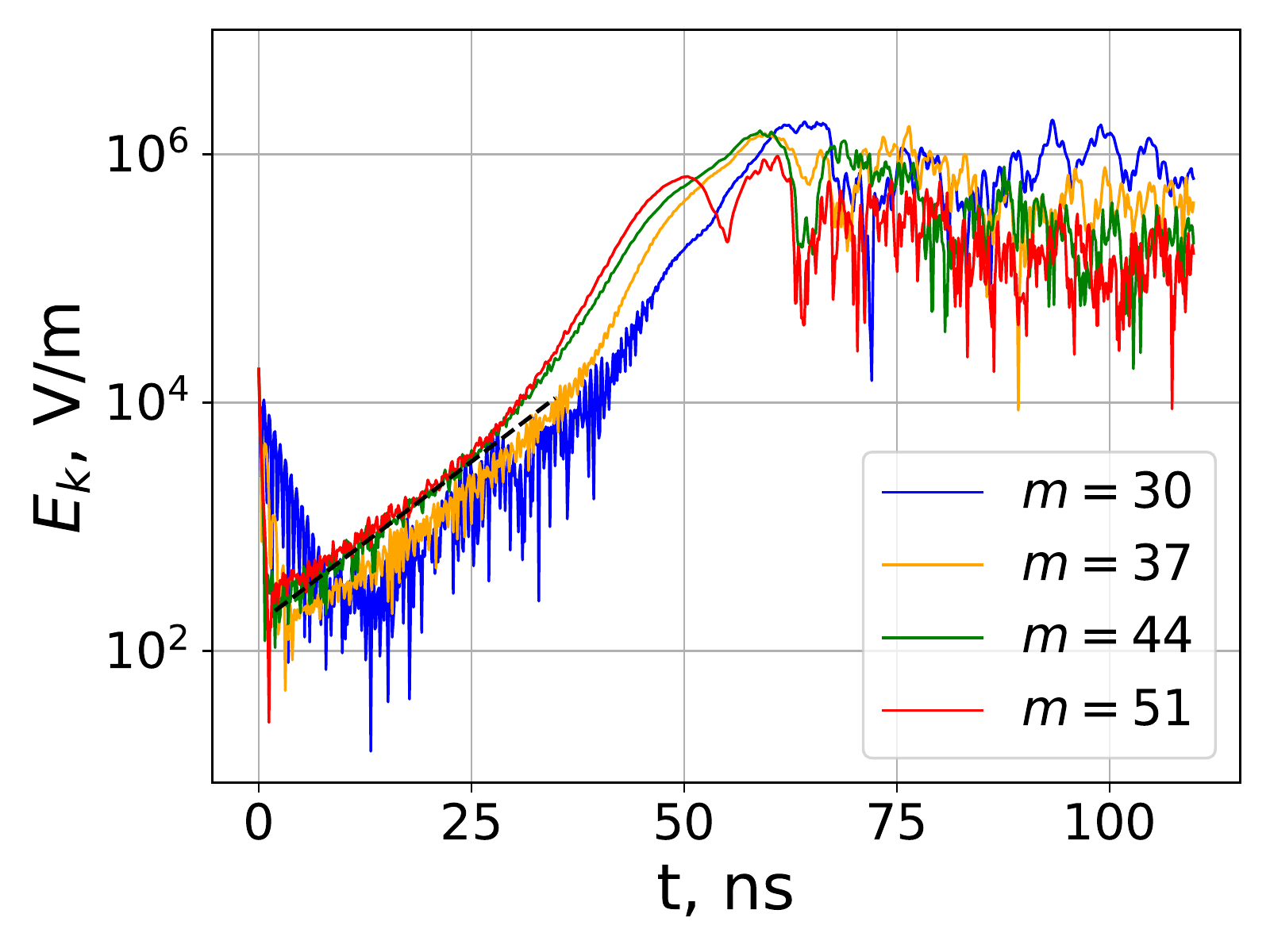}
\caption{The evolution of individual modes of the electric field, from PIC9 simulation. The dashed black line shows the fitted line on the $m=44$ mode.}
\label{V-qs-multi_M}
\end{figure}

In the PIC9 simulation, we have perturbed the group of the modes $m=\{30,37,44,51\}$. \Cref{table:growth_numbers_2vte_PIC9} shows that this method of initialization leads to a much smaller average standard error and indicates an improvement in the linearity of the growth compared to the PIC6, PIC7, and PIC8 simulations (see also \Cref{V-qs-multi_M}). This improved linearity, however, leads to a smaller 99\% confidence interval, and accordingly, the  theoretical average growth rate lies outside the 99\% confidence interval of the measurement. Nevertheless, the measured growth rate of PIC9 simulation are much closer to the theoretical values than those from the corresponding random-start simulation PIC2. 

\begin{table}[htbp]
\begin{tabular}[b]{|c|c|c|c|}
\hline
$m$  & \begin{tabular}[c]{@{}c@{}}$\gamma$ (Theory)\\ $\times 10^8s^{-1}$\end{tabular} & \begin{tabular}[c]{@{}c@{}}$\gamma$ (PIC9)\\ $\times 10^8s^{-1}$\end{tabular}&\begin{tabular}[c]{@{}c@{}}SE (PIC9)\\ \%\end{tabular} \\ \hline
30 & 0.90                                                                            & 1.32 &  3.63                \\ \hline
37 & 1.08                                                                            & 1.18 &  1.95                                                                   \\ \hline
44 & 1.17                                                                            & 1.26  &  1.23                                                                     \\ \hline
51 & 1.07                                                                            & 1.22 &  1.08                                                                     \\ \hline
\hline
Average & 1.06 & 1.25 & 1.97 \\ 
\hline
\end{tabular}
\caption{Comparison of the theoretical growth rates with the growth rates observed in the PIC9 simulation.\label{table:growth_numbers_2vte_PIC9}}
\end{table}

\begin{figure}[htbp]
\centering
\captionsetup[subfigure]{labelformat=empty}
\subcaptionbox{\label{EdiPic_100kmp_2vte_Ex_Growth-PIC10}}{\includegraphics[width=.49\linewidth]{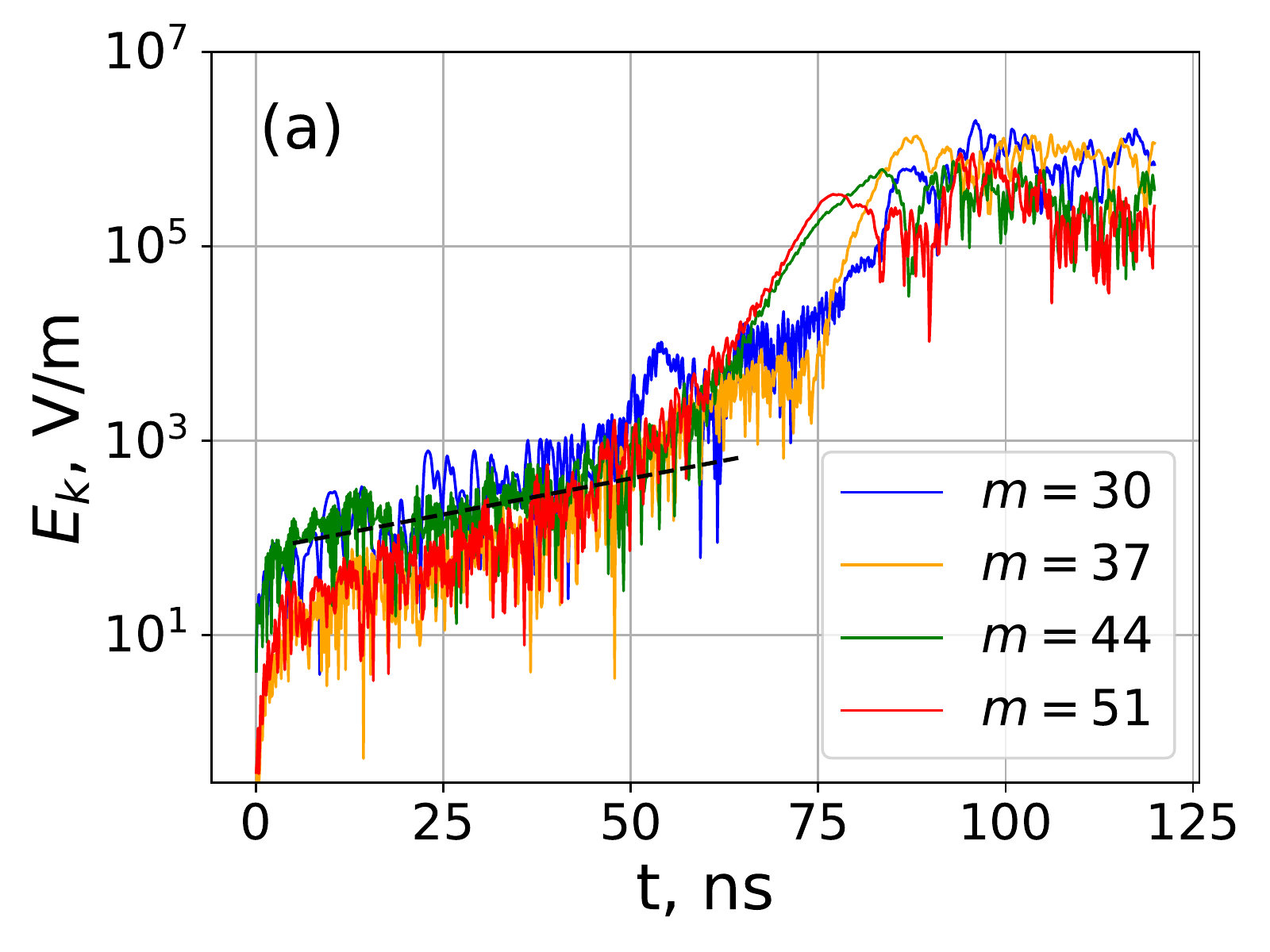}}
\subcaptionbox{\label{EdiPic_100kmp_2vte_Ex_Growth-PIC11}}{\includegraphics[width=.49\linewidth]{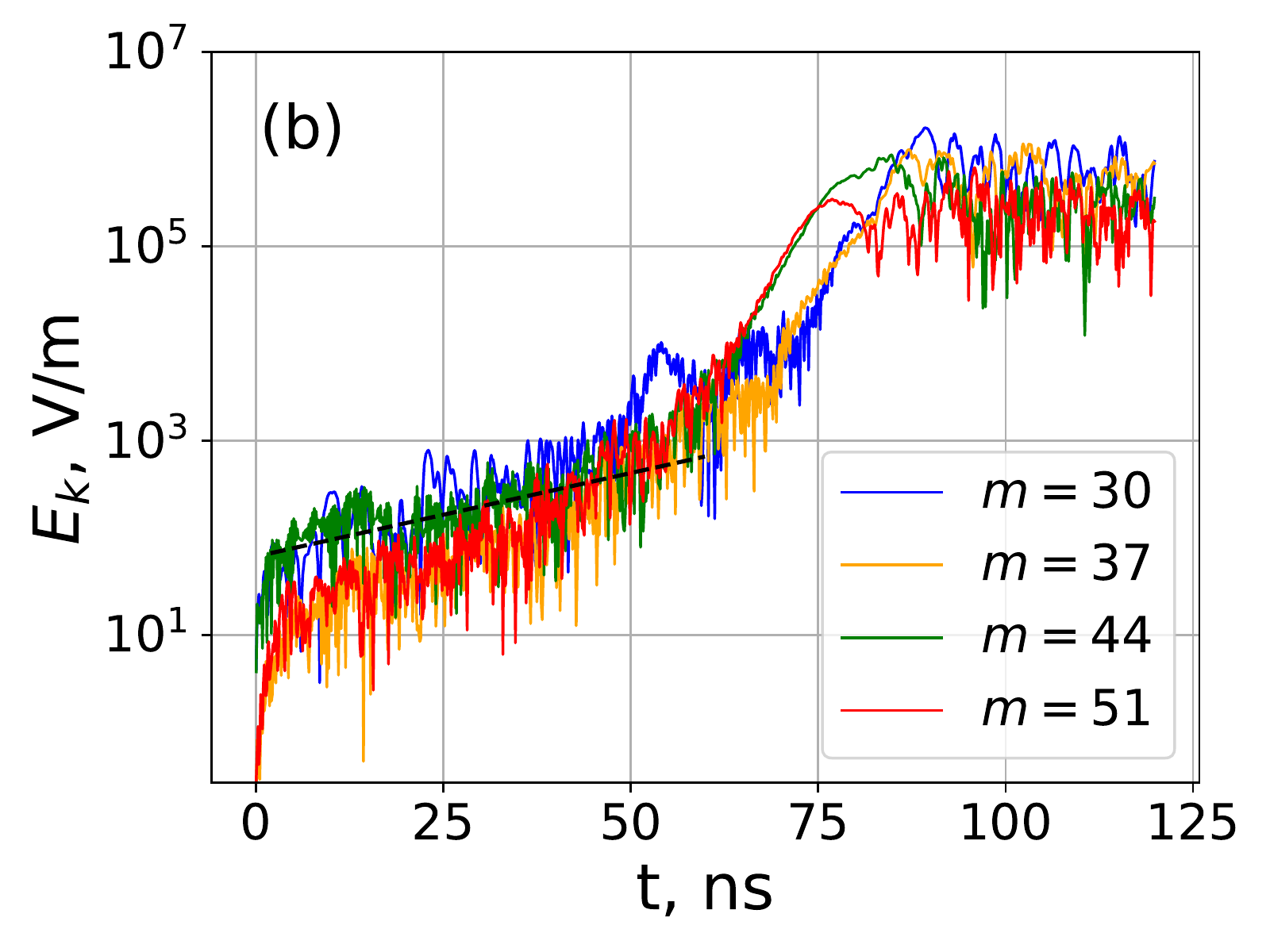}}

\subcaptionbox{\label{EdiPic_100kmp_2vte_Ex_Growth-PIC12}}{\includegraphics[width=.49\linewidth]{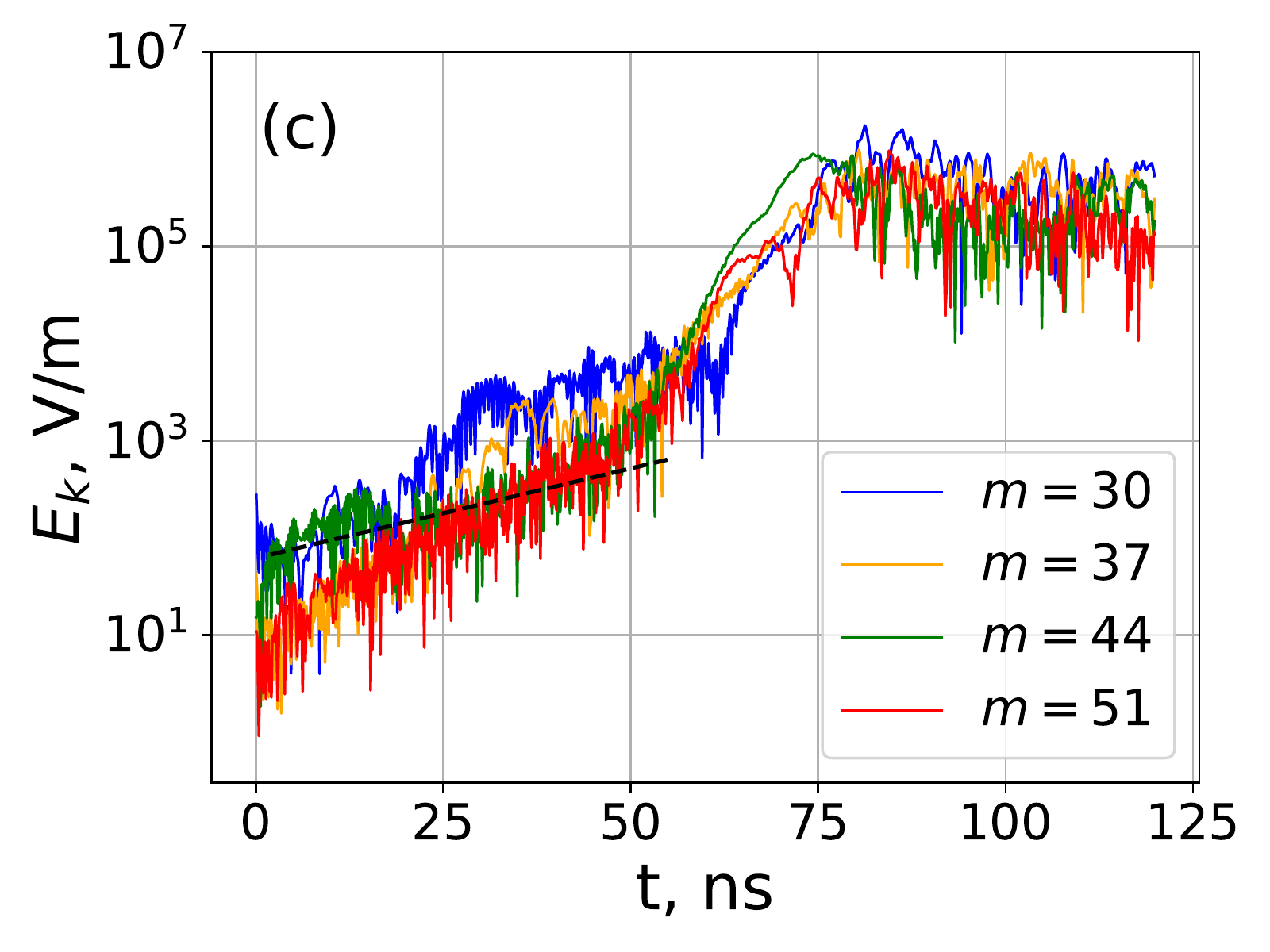}}
\caption{The evolution of individual modes of the  electric field in quiet-start PIC simulations using $10^4$ macroparticles per cell from a) PIC10  b) PIC11 c) PIC12 simulation. All figures are generated from the EDIPIC simulation code. The dashed black line shows the fitted line on the $m=44$ mode.}
\label{quiet_start_no44_Ek}
\end{figure}

To further investigate the role of the initial perturbation in the quiet-start simulations, we introduce three more PIC simulations (PIC10, PIC11, and PIC12). In these three simulations, the maximum growth rate $m=44$ is not perturbed initially. In the PIC10 simulation, we initialize the simulation with quiet-start but without exciting any mode. In this case, we expect the inherent noise of the PIC simulation to excite the instability. In PIC11 and PIC12, the simulations are initialized with perturbations in the modes $m=1$ and $m=31$, respectively. Although the individual modes start growing from much lower amplitudes than the random-start PIC simulations, their growth is quite oscillatory at the beginning (see \Cref{quiet_start_no44_Ek}). Also, the measured growth rates are mostly far from the theoretical values. However, in contrast to the random-start PIC simulations, most of the growth rates are underestimated by these simulations (see \Cref{table:growth_numbers_2vte_n044}). The average SNR of the growth rate is $-34.1$ dB in PIC10, $-34.3$ dB in PIC11, and $-30$ dB in PIC12. Therefore, the SNRs are lower than the quiet-start simulations PIC6, PIC7, and PIC8, where the maximum growth rate is perturbed initially.

\begin{table}[htbp]
\begin{tabular}[b]{|c|c|c|c|c|c|c|c|}
\hline
$m$  & \begin{tabular}[c]{@{}c@{}}$\gamma$ (Theory)\\ $\times 10^8$  s$^{-1}$\end{tabular} & \begin{tabular}[c]{@{}c@{}}$\gamma$ (PIC10)\\ $\times 10^8$ s$^{-1}$\end{tabular}& \begin{tabular}[c]{@{}c@{}}SE (PIC10)\\ \%\end{tabular} &\begin{tabular}[c]{@{}c@{}}$\gamma$ (PIC11)\\ $\times 10^8$ s$^{-1}$\end{tabular}& \begin{tabular}[c]{@{}c@{}}SE (PIC11)\\ \%\end{tabular}&\begin{tabular}[c]{@{}c@{}}$\gamma$ (PIC12)\\ $\times 10^8$ s$^{-1}$\end{tabular}& \begin{tabular}[c]{@{}c@{}}SE (PIC12)\\ \%\end{tabular} \\ \hline
30 & 0.90                                                                            & 0.62 & 4.19 & 0.62 & 4.14 & 1.21 &  2.59                                                               \\ \hline
37 & 1.08                                                                            & 0.59 & 6.11 & 0.61 & 5.67 & 1.26 &   1.52                                                                     \\ \hline
44 & 1.17                                                                            & 0.34 & 7.71 & 0.40 & 5.95 & 0.43 &   6.73                                                                     \\ \hline
51 & 1.07                                                                            & 0.63 & 4.52 & 0.65 & 4.88 & 0.92 &  3.13                                                                    \\ \hline
\hline
Average & 1.06 & 0.55 & 5.63 & 0.57 & 5.16 & 0.96 & 3.49 \\
\hline/
\end{tabular}
\caption{The comparison of the theoretical growth rates with the growth rates observed in PIC10, PIC11, and PIC12 simulations.\label{table:growth_numbers_2vte_n044}}
\end{table}

\section{Conclusion}\label{sec:conclusion}
In this study, we investigated the linear regime of the Buneman instability with several PIC and Vlasov simulations. The different PIC codes show good consistency of their results, and two different implementations of  Vlasov simulations are also consistent with each other; the results between the PIC and Vlasov simulations, however, differ significantly.  
We show that for a relatively small streaming velocity, $v_0=2v_{te}$, the random-start PIC simulations do not reproduce the theoretical linear growth rates, whereas the low-noise Vlasov simulations reproduce them quite accurately.  
 We show that the reason for the discrepancy is the discrete particle noise inherent to PIC simulations. This is demonstrated by initializing Vlasov simulations with the initial conditions of the random-start PIC simulations, which, in the latter case, show a discrepancy similar to PIC results.
 
This discrepancy is further confirmed by the study of the growth-rate sensitivity to a small flattening in the electron VDF. In \Cref{sec:small_flattening}, we show that a small flattening significantly  increases the linear growth rates. In random-start  PIC simulations, the flattening of the electron distribution function  occurs as a result of the early trapping of electrons in the noise-driven potential, as can be seen in \Cref{fe_8ns_halfx_xes1_rand_1e4}. In \Cref{sec:large_v0}, we show that for large streaming velocity, $v_0=6v_{te}$, the random-start PIC simulations can reproduce the linear growth rates with a reasonable accuracy. 
This limit is close to  the cold-plasma limit ($T_i\rightarrow 0$ and $T_e\rightarrow 0$), so that the effects of electron VDF are not important, and the linear growth rates are close to  their maximum cold-plasma values.    

The noise of the PIC simulations can be further reduced by increasing the number of macroparticles. We show that the results  up to $10^5$ macroparticles per cell (PIC4 and PIC5 simulations) will reduce the discrepancy, though the growth rates measured in PIC simulations remain far from the linear theory. Computational resource constraints made it impractical to increase the number of macroparticles per cell much beyond $10^5$; in principle, such an increase would increase the accuracy. On the other hand, the Vlasov simulations were able to reproduce the linear growth rates accurately within these constraints.
 
The effect of the noise resulting from the random sampling of the phase space with a limited number of macroparticles can be partially mitigated by the 
quiet-start\cite{dawson1983particle}. The accuracy of the growth rates from PIC simulations is greatly increased using the quiet-start initialization method but only if the maximum growth rate mode is perturbed initially. However, the noise is still significant in the linear growth of the quiet-start PIC simulations, and some of the growth rates remain inconsistent with their theoretical values. Therefore, the quiet-start method is not likely to completely remedy the problem of excessive noise in practice. 
  
In the PIC simulations of this study, the quiet-start method uses a bit-reversed sequence in the spatial subspace and a uniform sequence in the velocity subspace\cite{birdsall2004plasma}. These sequences are chosen because of their relative popularity and regularity, which greatly decreases the initial noise level. We also tried a bit-reversed sequence in velocity space, but the resulting growth rates were not as accurate as from the uniform sequence, and therefore, we did not report the results. In practice,  other sequences may give different accuracy; however, a systematic comparison of the various proposed sequences is beyond the scope of this study.

The issue of increased noise in PIC simulations is also related to a more general discussion as to what degree the PIC method, which works in between the exact Klimontovich equation and the asymptotic Vlasov equation, describes reality. In the PIC approach, the finite-sized charged clouds may still experience some binary interactions absent in the Vlasov equation but to some degree resembling Coulomb particle collisions \cite{scheiner2019particle,okuda1970collisions,palodhi2019counterstreaming,dupree1963kinetic,dawson1983particle}.    

For the simulations reported in this study, we have calculated the standard error associated with the measurement of the growth rates. For our purpose, the SE also provides a measure of deviation from linear growth, i.e., the extent to which the observed growth is linear as theory suggests. In the random-start PIC simulations, we observe highly oscillatory growth, and therefore, the SE is the largest for these simulations. However, the inaccuracy of the growth rates is so large that they fall outside the 99\% confidence interval of the theoretical growth rates. On the other hand, the growth of the low-noise Vlasov simulation is clearly linear, and therefore, the SE is much less than unity for them. Depending on the simulation code, the SE in the quiet-start PIC simulations can be larger or smaller than the SE in the random-start PIC simulations. Another quantity that we calculate for our simulations is the signal-to-noise ratio (SNR) during the growth of unstable modes. The SNR is largest for the Vlasov simulations, indicating the relatively low power of the noise carried in these simulations. In contrast to the SE, we show that the SNR in the quiet-start PIC simulations is less than that using random-start. 

Some modern methods of PIC simulation are proposed to reduce the noise level for a given number of macroparticles. Among these methods, the remapping and the delta-f methods have gained special attention recently. In the remapping method, the microparticles of the PIC simulation are frequently interpolated to a grid in phase space \cite{wang2011particle,myers20174th}. In this way, the noise level is decreased, but the computational cost is increased. In the delta-f method, the known part of the distribution function is separated from its variation (i.e.,~$f=\bar{f}+\delta f$, where $\bar{f}$ is the known distribution function and $\delta f$ is its variation)\cite{aydemir1994unified,brunner1999collisional,allfrey2003revised}. The existing macroparticles are only used to resolve the variation part instead of the distribution function, and therefore, the computational resources are allocated efficiently to reduce noise. However, to keep the noise level small, the condition $\delta f/\bar{f}\ll 1$ should be satisfied in the simulation. This condition is usually satisfied in the linear regime of instabilities (as in this study), but it may not quite be satisfied in the nonlinear regime if the distribution function significantly deviates from its initial shape.  It is expected that the remapping or delta-f method may ameliorate the noise problem of PIC simulations reported in this study. However, confirming this expectation would require other experiments that are beyond the scope of this study. 

\section*{Acknowledgments}
 This work was supported in part by the U.S. Air Force Office of Scientific Research FA9550-21-1-0031,  NSERC Canada, and by computational facilities of Compute Canada. A.S. acknowledges illuminating discussions with S. Janhunen.  
 
 \section*{Author Declarations}
 \subsection*{Conflict of interest}
 The authors have no conflicts to disclose.
 
  \section*{Data Availability Statement}
 The data that support the findings of this study are available from the corresponding author
upon reasonable request.

\bibliographystyle{apsrev4-1}
\bibliography{Refs.bib}
	
\end{document}